\documentclass[11pt]{article}

\usepackage{graphicx,amsmath,upgreek,bm}
\usepackage{colortbl}
\usepackage[rgb, table, dvipsnames]{xcolor}
\usepackage{booktabs}
\usepackage{makecell}
\usepackage{float}
\usepackage{pifont}
\usepackage{todonotes}
\usepackage[toc,page]{appendix}
\usepackage{multirow}
\usepackage{subcaption} 	
\usepackage{amsmath,amsfonts,amsthm,bm,mathtools}
\usepackage{lipsum}
\usepackage{geometry}
\usepackage[backend=bibtex, style=numeric-comp, sorting=none, maxbibnames=99]{biblatex}
\usepackage{algorithm}
\usepackage[noend]{algpseudocode}
\usepackage{arydshln}
\usepackage{quotes}
\definecolor{refcolor}{rgb}{0.65,0,0.35}
\usepackage[colorlinks=true]{hyperref}
\hypersetup{	
	linkcolor = refcolor,
	citecolor = refcolor,
	filecolor = refcolor,      
	urlcolor = blue,
	citecolor = refcolor
}

\graphicspath{ {}{../} }

\title{\bf Use of Topological Data Analysis for the Detection of Phenomenological Bifurcations in Stochastic Epidemiological Models}

\author{Sunia Tanweer$^{1,3}$\footnote{tanweer1@msu.edu} \and Konstantinos Mamis$^{2}$\footnote{kmamis@uw.edu} \and Firas A.~Khasawneh$^{3}$\footnote{khasawn3@msu.edu}}

\date{1. Dept. of Mechanical Engineering, Michigan State University \\ 2. Dept. of Applied Mathematics, University of Washington, Seattle, 98195, WA, USA \\ 3. Dept. of Computational Mathematics, Science and Engineering, Michigan State University}

\begin{document}
\maketitle

\vspace{-10mm}
\par\noindent\rule{\textwidth}{0.4pt}
\section*{Abstract}
\label{sec:abstract} 

We investigate predictions of stochastic compartmental models on the severity of disease outbreaks. The models we consider are the Susceptible-Infected-Susceptible (SIS) for bacterial infections, and the Susceptible -Infected-Removed (SIR) for airborne diseases. Stochasticity enters the compartmental models as random fluctuations of the contact rate, to account for uncertainties in the disease spread. We consider three types of noise to model the random fluctuations: the Gaussian white and Ornstein-Uhlenbeck noises, and the logarithmic Ornstein-Uhlenbeck (logOU). The advantages of logOU noise are its positivity and its ability to model the presence of superspreaders. We utilize homological bifurcation plots from Topological Data Analysis to automatically determine the shape of the long-time distributions of the number of infected for the SIS, and removed for the SIR model, over a range of basic reproduction numbers and relative noise intensities. LogOU noise results in distributions that stay close to the endemic deterministic equilibrium even for high noise intensities. For low reproduction rates and increasing intensity, the distribution peak shifts towards zero, that is, disease eradication, for all three noises; for logOU noise the shift is the slowest. Our study underlines the sensitivity of model predictions to the type of noise considered in contact rate. 
\vspace{0mm}

\section{Introduction}
\label{sec:intro}

Accurate predictions of the spread of communicable diseases are a valuable tool for policy making during disease outbreaks, such as the recent COVID-19 pandemic, as well as for endemic diseases, such as gonorrhoea. Since their inception \cite{Kermack1927}, compartmental models have been a simple yet efficient tool for the quantification of disease spread in a population \cite{Capasso1978, Hethcote1994, Brauer2008, Capasso1993, Diekmann2021}.

In classical compartmental models, the total population is divided into disjoint states with respect to the disease, e.g. susceptibles or infected, with the progression of individuals through the states being determined by parameters such as the transmission or curing rates. For the deterministic case, the model parameters are considered constants, calculated as averages over the total population; the resulting model is a system of ordinary differential equations that has as state variables the number of individuals in each compartment. Determining the equilibrium points of the deterministic compartmental models and performing bifurcation analysis results in conditions for the persistence or eradication of the disease from the population, as well as estimates for the number of infected or removed individuals in the long time, which is a measure of the severity of the disease outbreak. 

However, the parameters in compartmental models are subject to random fluctuations originating from environmental factors, variability in human behavioral patterns, and also changes in the pathogen itself~\cite{Allen2017}. Incorporation of random fluctuations in model parameters gives rise to stochastic compartmental models in the form of stochastic differential equations (SDEs).  The stochastic counterpart of bifurcation analysis of the equilibrium points in deterministic models is the detection of changes in the shape of the long-time probability density function (PDF) or the Kernel Density Estimate\footnote{KDEs provide a non-parametric way to generate a smooth estimate of the underlying PDF of a random variable by averaging kernel functions (typically a gaussian) centered at the observed data points.} (KDE) for the infected or removed fraction of the population. Such changes are called phenomenological (P-)bifurcations~\cite{NIT}. 

In stochastic compartmental models, the most common choice is to incorporate random fluctuations by perturbing the model parameters that are volatile, such as the contact rate, by Gaussian white noise \cite{Gray2011, Ji2014, Cai2015, Meng2016, Cai2020}. However, it has recently been shown that white noise perturbation results in a systematic underestimation of the disease outbreak \cite{Mamis2023, Mamis2024}. Additionally, from a modeling perspective, white noise cannot take into account the correlations stemming from the pattern of human social interactions that greatly affect the transmission of disease. Instead, correlated Ornstein-Uhlenbeck (OU) noise has been proposed for parameter perturbation in biological systems \cite{Liu2022, Wang2018, Bartoszek2017, Rohlfs2013, Aalen2004}, since OU noise combines the modeling of stochastic fluctuations with the stabilization around an equilibrium point, due to its mean-reverting property \cite{Allen2016}. Furthermore, Ornstein-Uhlenbeck perturbations in contact rate do not result in an underestimate of the severity of the disease outbreak, with results that stay close to observed data, for example, for the Omicron wave of the COVID-19 pandemic \cite{Mamis2023, Mamis2024}. 

The main objection to the use of white or Ornstein-Uhlenbeck noises for the perturbation of the strictly positive contact rate is that they may take negative values, since they are unbounded Gaussian processes. Previous work in stochastic oncology remedies these unwanted negative values in stochastically perturbed parameters by considering bounded noise \cite{Domingo2019, dOnofrio2010}; however, noise-induced transitions in compartmental models under bounded noise have not yet been studied extensively \cite{Bobryk2005}. 

For compartmental models, logOU noise, that is, a noise whose logarithm is an OU process, has recently been proposed for the perturbation of the contact rate. LogOU noise is an adequate choice due to its positive only values \cite{Liu2024, Shi2023}, and its ability to model the presence of superspreaders in the population, by its long distribution tail \cite{Faranda2020, SETC2020}. The mathematical results of the aforementioned works have been on the derivation of conditions for the eradication or persistence of a disease in the population under logOU noise. In the present work, we determine how logOU noise perturbations affect the final size of the disease outbreak, by determining the number and locations of the peaks in the stationary probability density of the infected or removed, and how they change as the noise intensity or model parameters vary. 

In the case of uncorrelated Gaussian white noise perturbations, the stationary probability density of the infected or removed is determined as the stationary solution of the corresponding Fokker-Planck equation~\cite{Gardiner2004}. For stochastic systems under correlated Gaussian noise, nonlinear Fokker-Planck equations that are accurate even for large correlation times have been formulated~\cite{Mamis2021, Mamis2019}; recently, their stationary solutions have been used to determine the stationary probability density of the infected in models for bacterial disease transmission under Ornstein-Uhlenbeck noise \cite{Mamis2023, Mamis2024}. 

Unfortunately, since the existing Fokker-Planck formalism only applies to SDEs under Gaussian noise, it cannot be used for epidemiological models under logOU noise. This, along with the hurdle of visual inspection of the stationary solutions, is a key challenge in the bifurcation analysis of these SDEs. For this reason, in the present work, we employ the recently devised \textit{homological bifurcation plots}~\cite{Tanweer2024, Tanweer2024b} based on direct Monte Carlo simulations of the SDEs. Homological bifurcation plots offer a powerful topological lens that excel at detection of P-bifurcations where traditional methods fall short. These plots are built upon the cubical persistence framework in topology which allows capturing the qualitative shape of PDFs by quantifying the number of peaks, limit cycles and higher dimensional structures. By capturing such qualitative changes in the probability distributions, these plots provide a robust, noise-independent framework which is effective under all noise conditions. Their ability to identify structural changes in the distributions without requiring the analytical expressions makes them particularly valuable for complex epidemiological models, where direct visual inspection becomes infeasible across large parameter spaces. Moreover, in multimodal distributions, these plots also give information about the relative heights of the modes, that is, the height difference between a peak and the saddle point connecting it to another peak. In addition to homological bifurcation plots, we introduce complementary \textit{peak-tracking} plots, which track the precise location of the highest likelihood across the changing parameter space, not merely the number of peaks. This visualization method provides crucial quantitative information about how the most probable states shift and/or drift as system parameters are varied. These peak-tracking plots capture the evolution of the system's most likely states, both pre- and post-bifurcation. Together, these plots provide a thorough characterization of the bifurcation landscape in SDEs. In this manuscript, we explore the impact of different types of noise on stochastic SIS and SIR models, and use these plots for finding noise-induced transitions in these models over a large parameter space, providing insights into the spread of disease under uncertainty. 

\section{Mathematical Background}
\label{sec:math-background}
\subsection{Compartmental Models in Epidemiology}
We consider two fundamental compartmental models, from which more detailed models are derived: the Susceptible-Infected-Susceptible (SIS), and the Susceptible-Infected-Removed (SIR) model. We then present the different noises for random fluctuations that we will investigate in the present work, namely the Gaussian white noise, the Ornstein-Uhlenbeck noise, and the logarithmic Ornstein-Uhlenbeck noise.
\subsubsection{Deterministic Compartmental Models}
\begin{figure}[!htbp]
    \centering
    \includegraphics[width = 0.4\linewidth]{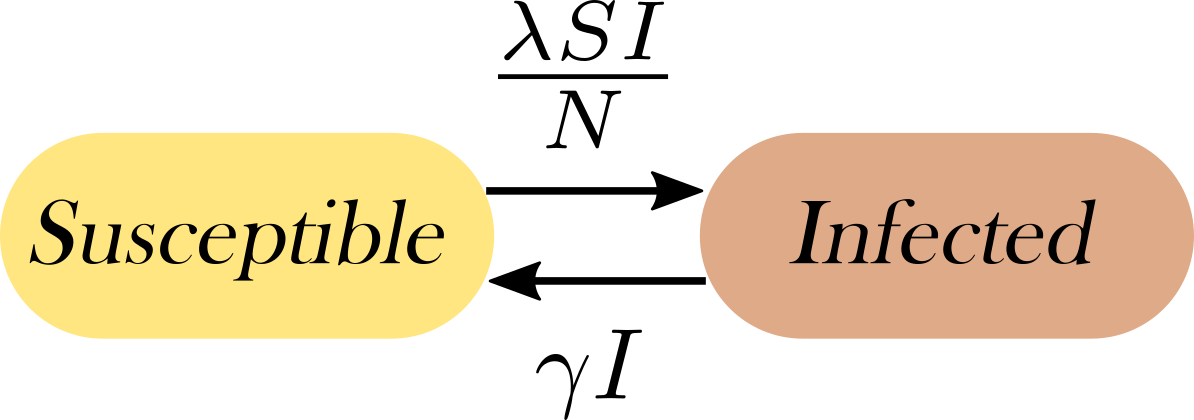}    \caption{Susceptible-Infected-Susceptible model with parameters $\lambda$ and $\gamma$ for the contact and curing rates respectively.}
    \label{fig:SIS}
\end{figure}

The SIS model is suitable for diseases whose previous infection does not confer immunity; an individual can be infected immediately after recovery, as is the case for most bacterial and sexually transmitted diseases~\cite{KosekDygas1988, stochNUM}. The population $N$ is divided into Susceptibles $S(t)$, who become infected with a rate $\lambda$, and Infected $I(t)$ who recover with a rate $\gamma$ (Fig.~\ref{fig:SIS})
\begin{align}\label{eq:sis}
    \frac{dS}{dt} &= -\lambda \frac{SI}{N} + \gamma I \\
    \frac{dI}{dt} &= \lambda \frac{SI}{N} - \gamma I
\end{align}

Since demographic dynamics are omitted, population $N$ is considered constant. Normalizing the number of infected by the total population, we re-write the SIS model as
\begin{equation}\label{eq:sis_normalized}
    \frac{di}{dt}=\lambda i(1-i)-\gamma i,
\end{equation}
where $i(t)\in[0,1]$ is the infected fraction of the population. In a deterministic SIS model, the stability analysis is straightforward: For basic reproduction number $R_0=\lambda/\gamma<1$, the SIS model has its only stable equilibrium at zero, corresponding to disease eradication. For $R_0>1$, the disease persists in the population and the stationary value $1-R_0^{-1}$ of the infected population fraction is a measure of the disease outbreak. 

\begin{figure}[!htbp]
\centering
\includegraphics[width = 0.65\linewidth]{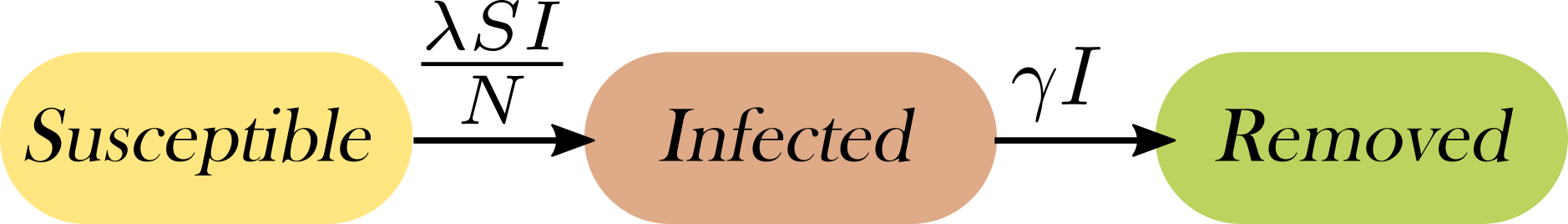}
\caption{Susceptible-Infected-Removed (SIR) model with parameters $\lambda$ and $\gamma$ for the contact and the curing rates respectively.}
\label{fig:SIR}
\end{figure}

Diseases such as seasonal flu that confer immunity are best described by the SIR model; the additional compartment $R$ of Removed contains individuals that no longer participate in disease transmission, either due to immunity or death after infection. The progression of individuals through the three compartments (Fig.~\ref{fig:SIR}) is modeled by the equations: 
\begin{align}
    \frac{dS}{dt} &= -\lambda \frac{SI}{N} \\
    \frac{dI}{dt} &= \lambda \frac{SI}{N} - \gamma I\\
    \frac{dR}{dt} &= \gamma I.
\end{align}

Under the assumption $S(t)+I(t)+R(t)=N$ of constant population, the system of equations is reduced to two
\begin{align}
    \frac{di}{dt} &= \lambda i(1-i-r) - \gamma i\label{eq:i}\\
    \frac{dr}{dt} &= \gamma i,
\end{align}
where $i(t)=I(t)/N\in[0,1]$ and $r(t)=R(t)/N\in[0,1]$ are the infected and removed population fractions respectively. Unlike the SIS model, the behavior of the SIR model depends on the initial populations too: From Eq.~\eqref{eq:i}, we observe that, in order for the initial number of infected to increase, $R_0s_0>0$, where $R_0=\lambda/
\gamma$ is the basic reproduction number and $s_0$ is the initial number of susceptibles. Another difference between SIS and SIR models is that, in the SIR, the infected population is transient, since its long-time value is zero. Thus, the severity of the outbreak of a SIR disease is described by the long-time value of the removed, which is proportional to the cumulative number of infections.

\subsubsection{Stochastic Compartmental Models}
From the parameters in the SIS and SIR models, contact rate $\lambda$ is the most volatile, since curing rate $\gamma$ depends largely on the pathogen's biology. To account for random fluctuations in $\lambda$, we study three options. First, we consider an additive perturbation by a zero-mean Gaussian noise (white or OU) of intensity $\sigma$ around the mean value $\bar{\lambda}$:
    \begin{equation}\label{eq:lambda_additive}
        \lambda(t)=\bar{\lambda}+\sigma\xi(t)
    \end{equation}
For the case of $\xi(t)$ being white noise, the autocovariance function of contact rate reads
\begin{equation}\label{cov:white}
     \mathrm{Cov}\left[\lambda^{WN}(t_1)\lambda^{WN}(t_2)\right]=\sigma^2\delta(t_1-t_2),
\end{equation}
with $\delta(t_1-t_2)$ being Dirac's delta function. For the case of $\xi(t)$ being OU noise, the autocorrelation function reads
\begin{equation}\label{cov:OU}
   \mathrm{Cov}\left[\lambda^{OU}(t_1)\lambda^{OU}(t_2)\right]=\frac{\sigma^2}{2\tau}e^{-|t_1-t_2|/\tau},
\end{equation}
with $\tau>0$ being the correlation time of the process. Note that for $\tau\rightarrow0$, OU autocovariance \eqref{cov:OU} results in the white noise autocovariance \eqref{cov:white}. 

As an alternative to Gaussian noise that generates strictly positive fluctuations, we consider the positive, logarithmic OU noise. In this case, we consider the multiplicative perturbation by the exponential of a zero-mean OU noise $\eta(t)$ with intensity $D$ and correlation time $s$ around the median value $\bar{\lambda}$:

\begin{align}
    \lambda^{\log OU}(t) = \bar{\lambda}\exp(\eta(t)),
\end{align}
and thus, contact rate is distributed as
\begin{equation}\label{lambda_OU_dist}
    \lambda^{\log OU}(t)\sim\mathrm{LogNormal}\left(\log(\bar\lambda),\frac{D^2}{2s}\right),
\end{equation}
while we determine its autocovariance to
\begin{equation}\label{auto_logOU}
    C(t_1-t_2)=\mathrm{Cov}\left[\lambda^{\log OU}(t_1)\lambda^{\log OU}(t_2)\right]=\bar\lambda^2\exp\left(\frac{D^2}{2s}\right)\left[\exp\left(\frac{D^2}{2s}e^{-|t_1-t_2|/s}\right)-1\right].
\end{equation}
From Eq.~\eqref{auto_logOU}, we calculate the correlation time of $\lambda^{\log OU}(t)$ in terms of parameters $D$, $s$:
\begin{equation}\label{tau_logOU}
    \tau^{cor}_{\log OU}=\frac{\int_0^{\infty}C(u)du}{C(0)}=s\frac{\mathrm{Ei}\left(\frac{D^2}{2s}\right)-\log\left(\frac{D^2}{2s}\right)-\gamma_e}{\exp\left(\frac{D^2}{2s}\right)-1},
\end{equation}
where $\mathrm{Ei}(\cdot)$ is the exponential integral function and $\gamma_e$ is the Euler–Mascheroni constant.
\subsubsection{Comparison of results under different random fluctuations}
\label{subsubsec:comparison_OU_logOU}
\begin{figure}[!htbp]
\centering
\includegraphics[width=0.3\linewidth]{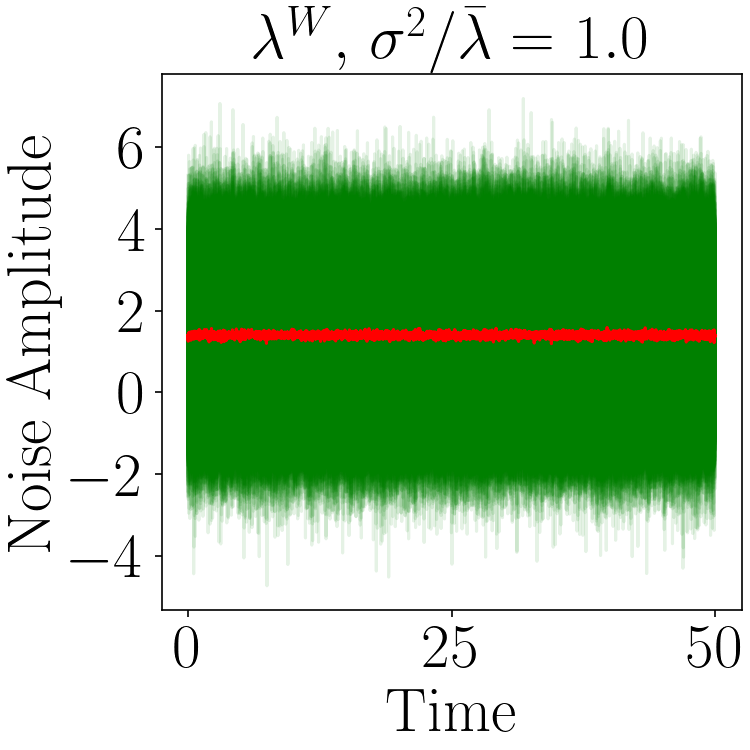}
\includegraphics[width=0.3\linewidth]{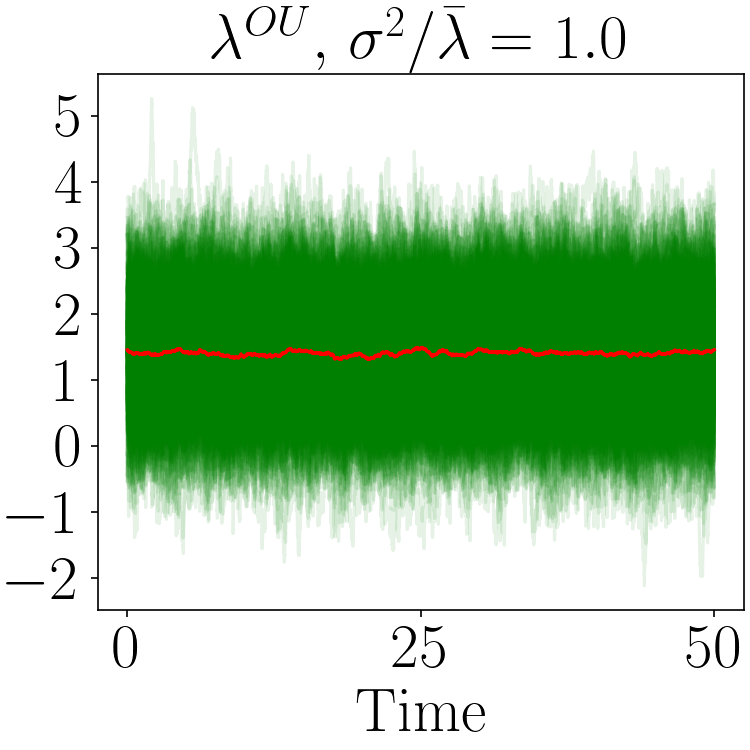}
\includegraphics[width=0.3\linewidth]{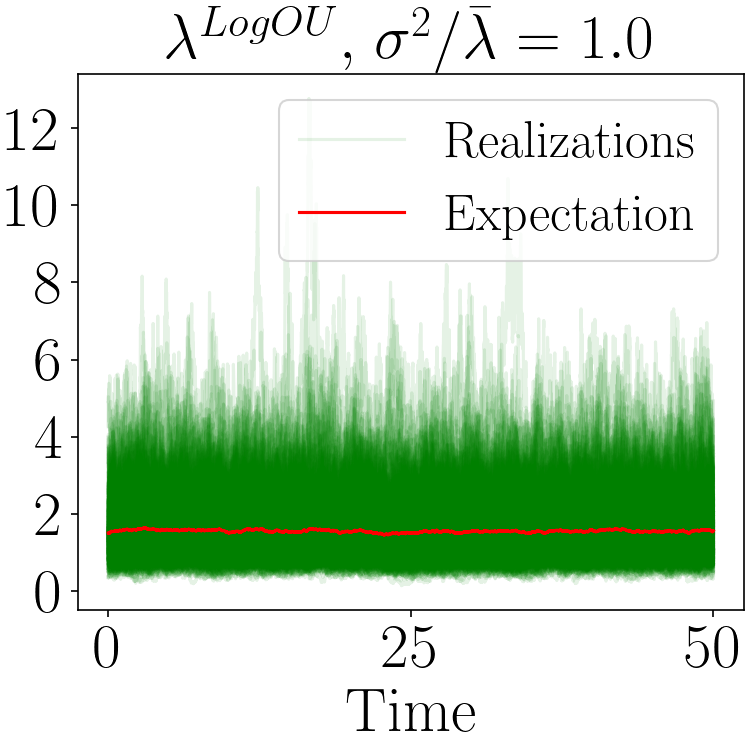}
\\
\includegraphics[width=0.3\linewidth]{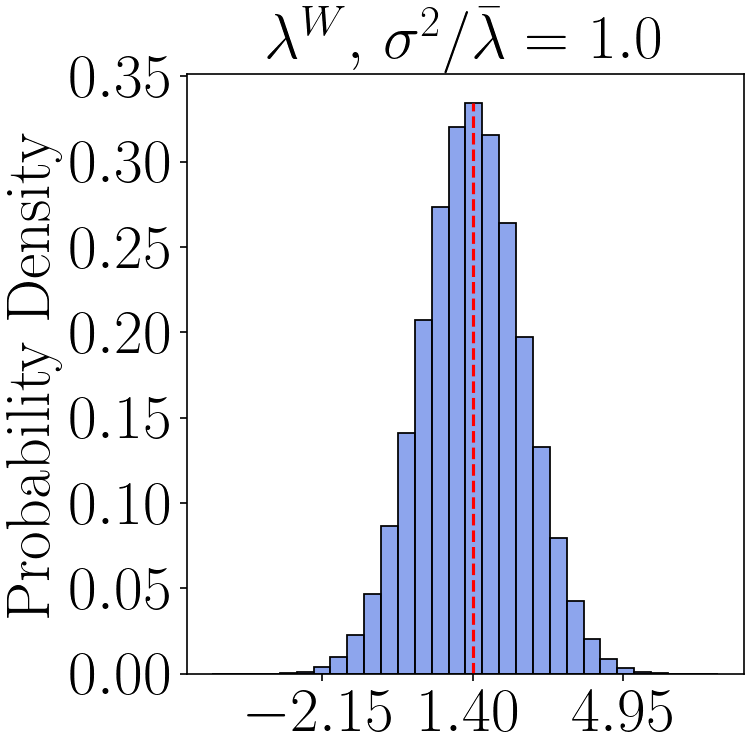}
\includegraphics[width=0.3\linewidth]{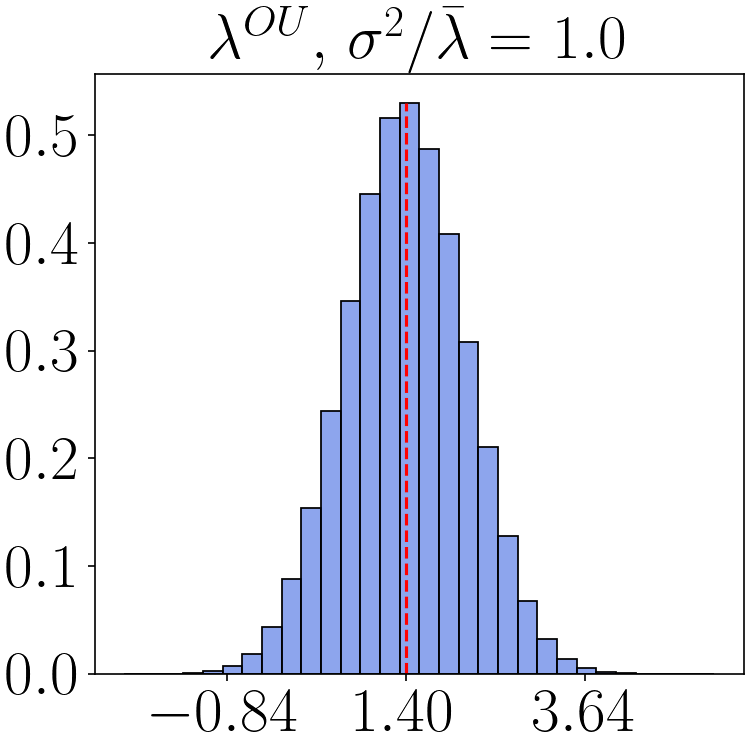}
\includegraphics[width=0.3\linewidth]{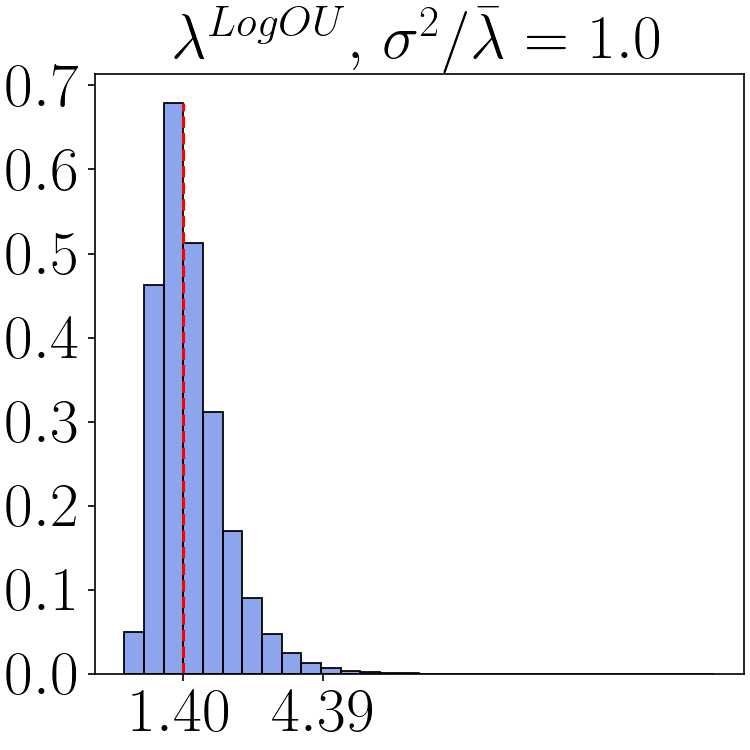}
\caption{Trajectories for contact rate $\lambda$ having $\bar{\lambda} = 1.4 \text{ per month}$ with white, OU and logarithmic OU noise perturbations, followed by the distributions for each---for a relative noise intensity of 1.0. Basic reproduction number $R_0 = 1.4$ (gonorrhea) was taken for these simulations.}
\label{fig:noise_representations_1}
\end{figure}

The counterpart of equilibrium points for stochastic models are the stationary probability density functions (PDF) of their response. More specifically, the stationary PDF of the infected for SIS, and of the removed fraction for SIR quantify the severity of the disease outbreak that the stochastic compartmental models predict. Thus, we are interested in determining the stationary PDFs mentioned above for a range of the model parameters, as well as of the parameters of the different noises (white, OU, logOU) considered (Fig.~\ref{fig:noise_representations_1}); we are interested in determining whether PDF is narrow or broad, and where its peaks are located, since they indicate high concentrations of probability mass and thus the most likely values for the infected or removed population fractions. 

In our recent work \cite{Mamis2024}, we analytically derived that, in an SIS model and for Gaussian noise perturbations in contact rate, the dimensionless parameters that determine the stationary PDF of the infected population fraction are: the basic reproduction number $R_0$, the relative noise intensity $\sigma^2/\bar\lambda$, and for the case of OU noise, the relative correlation time $a=\tau(\lambda-\gamma)$, with $(\lambda-\gamma)^{-1}$ being a characteristic timescale of the SIS model. To the best of our knowledge, there are no analytic results for the stationary PDF of compartmental models under logarithmic Ornstein-Uhlenbeck noise; in the present work, we rely on simulations. 

\begin{figure}[!htbp]
\centering
\includegraphics[width=0.3\linewidth]{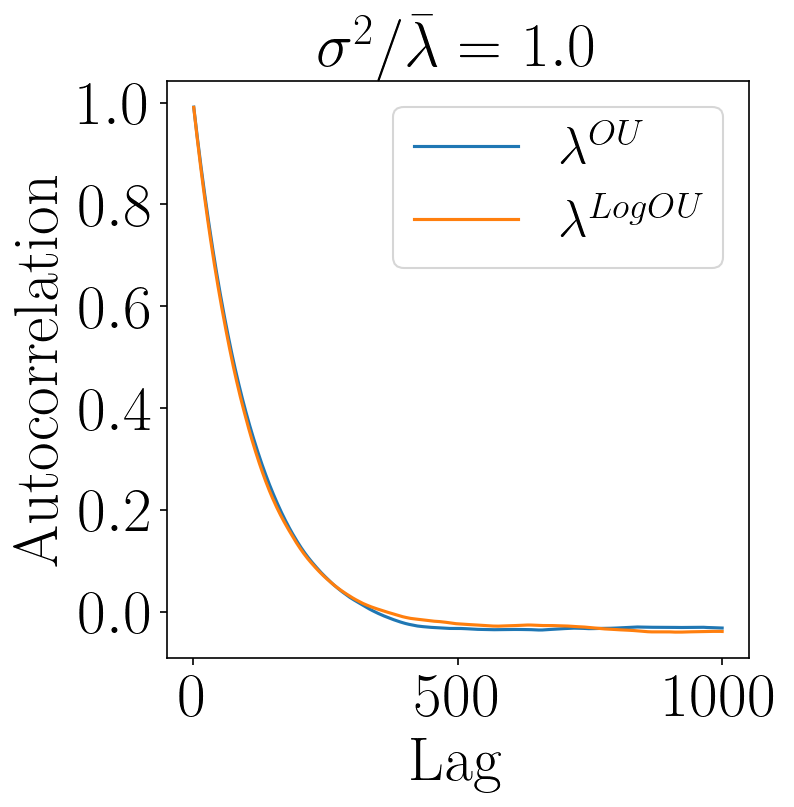}
\caption{Autocorrelation for contact rate $\lambda$ having $\bar{\lambda} = 1.4 \text{ per month}$ with OU and logarithmic OU noise perturbations---for a relative noise intensity of 1.0. Basic reproduction number $R_0 = 1.4$ (gonorrhea) was taken for these simulations.}
\label{fig:autocorrelation_1}
\end{figure}

For comparison of results, it is important to ensure that the noise parameters of intensity and correlation time are similar for the different noises considered.
To ensure this, we determine the parameters $D$ and $s$ for the logOU noise in Eq.~\eqref{lambda_OU_dist} by setting the variance and correlation time of the logOU equal to those of the OU noise to which we want to compare; this is performed by using the Eqs.~\eqref{auto_logOU} and~\eqref{tau_logOU}. See Fig.~\ref{fig:autocorrelation_1} for the empirically computed autocorrelation for both noises.

\subsection{Topological Data Analysis (TDA)}
One key topological space we need is the cubical complex, which is helpful for discretized function data in matrix representation. The sections below provide a general outline of homology and superlevel-set persistent homology---referring the interested reader to full details about works by~\cite{Dey2022, Oudot2015}.

\subsubsection{Cubical Complex}
The cubical complex is constructed from simple intervals: either a unit length interval, $[u, u+1]$, or a degenerate interval, $[u,u]=[u]$ where $u$ is an element of the set of integers $\mathbb{Z}$. A cube in an $n$-dimensional space is: $I = \prod_{i = 1}^{K}[u_i, u'_i]$ where $u'_i$ is either $u_i$ or $u_{i+1}$. The dimension of the resulting cube is given by the number of non-degenerate intervals in the above product. By way of example, a $0$-cube is a vertex, a $1$-cube an edge, a $2$-cube a square, and a $3$-cube a voxel, and so on. Note that the coefficients are in $\mathbb{Z}_2$, hence the orientation of the cubes is not considered.

If one cube $\sigma$ is a subset of another $\tau$, we say that $\sigma$ is a face of $\tau$, we write $\sigma \leq \tau$. For example, both edges $\sigma_1 = [0] \times [0,1]$ and $\sigma_2 = [0,1] \times [0]$ are faces of the square $\tau = [0,1]\times [0,1]$. Thus, a cubical complex is a set of cubes such that the requirement for including a $d$-cube in the complex is to include all its faces.

\subsubsection{Homology}
In algebraic topology, homology serves as a mathematical tool for understanding the structure of spaces by quantifying and describing key elements like connected parts, loops, and empty regions. For our discussion, let's consider a specific fixed space denoted as $X$, typically organized as a structured pattern of simplices or cubes.

The fundamental concept behind homology revolves around examining building blocks of various dimensions—depicted as simplices or cubes—and comprehending their interconnections. These building blocks act as the fundamental constituents forming the space. The $p$th chain group, $C_p(X)$, represents combinations of $p$-dimensional building blocks, while the boundary map $\partial_p$ illustrates how these blocks link together.

Homology $H_p(X)$ is defined as the collection of cycles considering boundaries. A cycle signifies a combination of building blocks forming a closed loop, while a boundary represents a combination that lies on the perimeter of something. Consequently, the collection of $H_p(X)$ for different values of $p$ informs us about the distinct loops, voids, or connected parts within the $X$. 

Mathematically, this concept is expressed as $H_p(X) = \text{Ker}(\partial_p) / \text{Im}(\partial_{p+1})$, where $\text{Ker}(\partial_p)$ denotes the set of cycles (combinations forming closed structures), and $\text{Im}(\partial_{p+1})$ represents the set of boundaries (combinations situated on the edge). Thus, the quotient, $H_p(X)$, captures the unique and non-duplicative aspects of the space at the $p$-dimensional level. Informally, the homology of a space provides a vector describing topological characteristics inherent to the space with each homology dimension $p$ providing some aspect of the shape of the underlying space such as connected components ($p=0$), loops ($p=1$), voids ($p=2$), or higher-dimensional counterparts for $p>2$. 

\subsubsection{Superlevel Persistent Homology}

Persistent homology is a variant of homology and one of the most common tools associated with TDA. It encodes information about the structure of a parameterized space by following the homology changes of the latter as a filtration parameter changes~\cite{Dey2022}.

Let $X$ be a topological space with a function $f:X \to \mathbb{R}$. Let data be given as a set of real values $a_1\leq a_2 \leq \cdots \leq a_n$. We can now introduce two types of filtrations, denoted as the \textit{sub-level-set} and the \textit{super-level-set}. The first one gives us $X_{a_1} \subseteq X_{a_2} \subseteq \cdots \subseteq X_{a_n}$ where $X_a = f^{-1}(-\infty,a]$. The latter produces $X^a = f^{-1}[a,\infty)$ which results in $X^{a_n} \subseteq X^{a_{n-1}} \subseteq \cdots \subseteq X^{a_1}$.

A fundamental theorem of algebraic topology says that homology is functorial. In the context of inclusions, this means that there is an induced map between the relevant homology groups. With these we define both a sub-level-set and a superlevel-set homology filtration, progressing from $H_p(X_{a_1})$ to $H_p(X_{a_n})$ and $H_p(X^{a_n})$ to $H_p(X^{a_1})$, respectively. 

\begin{figure}[!htbp]
    \centering
    \includegraphics[height = 1.5in]{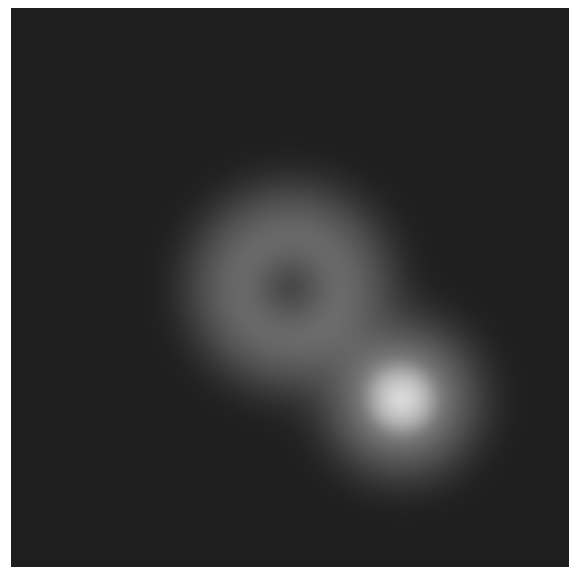} \includegraphics[height = 1.5in]{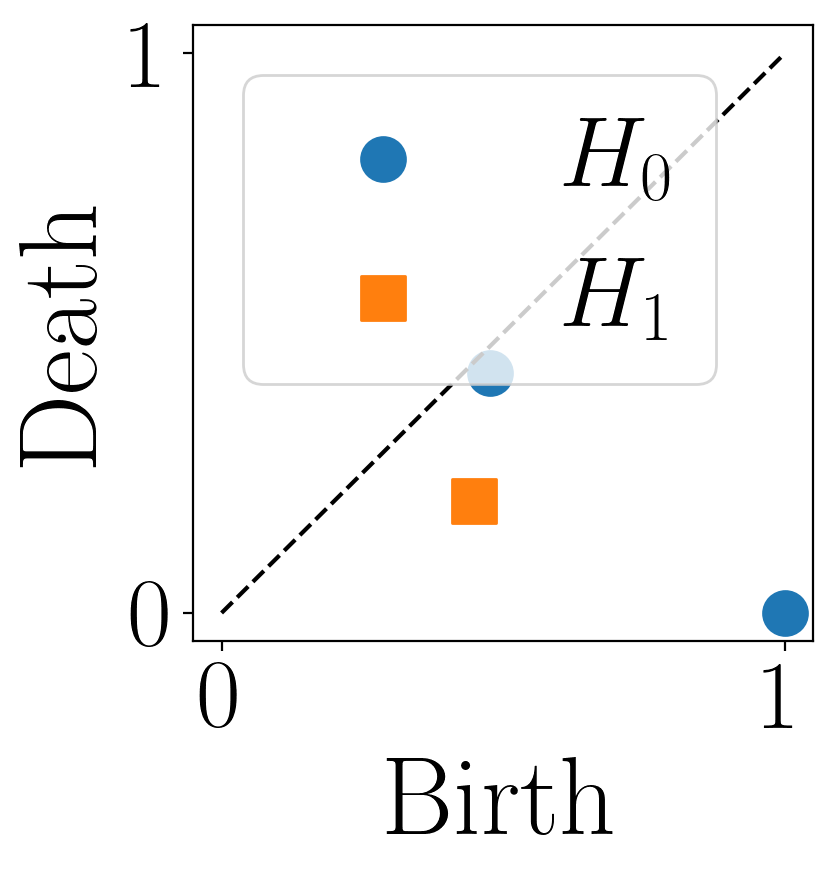}\\
    \includegraphics[height = 0.75in]{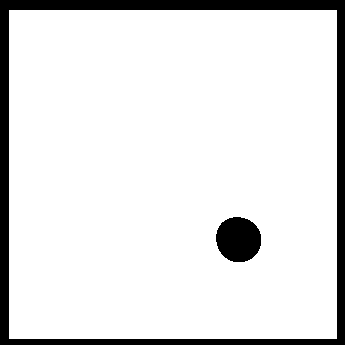}
    \includegraphics[height = 0.75in]{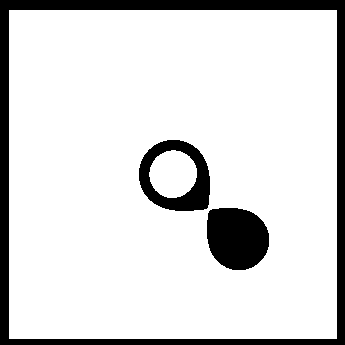}
    \includegraphics[height = 0.75in]{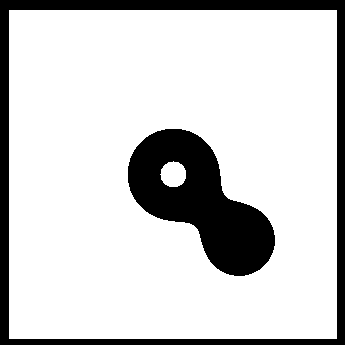}
    \includegraphics[height = 0.75in]{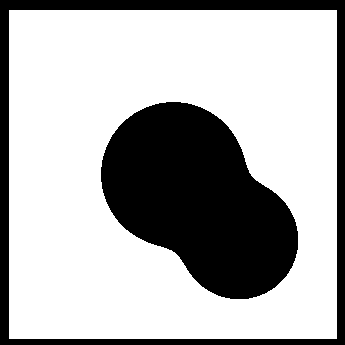}
    \caption{Superlevel cubical persistence of unit-normalized image data shown. Figures correspond to (a) $K^{0.8}$, (b) $K^{0.5}$, (c) $K^{0.4}$ and (d) $K^{0.2}$. Figures adapted from~\cite{Tanweer2024, Tanweer2024b} with modifications.}
    \label{fig:CubicalPersistence}
\end{figure}

Persistent homology further postulates that such a filtration can always be uniquely decomposed into pairs of $(b, d)$. In this regard, $b = a_i$ represents the birth of a new homology class in $H_p(X_{a_i})$ or $H_p(X^{a_i})$, which is different from the image of $H_p(X_{a_{i-1}})$ or $H_p(X^{a_{i+1}})$. $d = a_j$ is the point where it merges with the image of an older class entering $H_p(X_{a_j})$ or $H_p(X^{a_j})$.

This information conventionally appears in a persistence diagram displaying the pairs as dots in a plane. In particular, in super-level-set persistence, $b>d$, where $b$ is the value of the parameter when a structure first appears (several authors use the term "birth") and $d$ stands for its disappearance ("death"). As a result, superlevel-set points lie below the diagonal line $\Delta = {(x,x) \mid x \in \mathbb{R}}$.

In this paper, we are only interested in using one persistence: the super-level-set persistence of cubical complexes. Consider for this purpose a greyscale image or an $m \times n$ matrix $M$, defining on this its domain, the cubical set $K=\mathcal{K}([0,m]\times[0,n])$. Then the image is a function $f:K \to \mathbb{R}$ defined on that cubical set, where for any cube $s_{i,j}=[i,i+1]\times[j,j+1]$ from $K$, $f(s_{i,j})$ is the value of the matrix entry $M_{i,j}$. The super-level-set of $f$, defined as $K^a = f^{-1}[a, \infty)$, is also a cubical complex. Studying the super-level-set persistence of this map, we can study its evolution as the parameter (in this case, $a$) changes, and along with it, different cubical complexes and their topological features evolve. Fig.~\ref{fig:CubicalPersistence} shows an example of image data and evolution of homology: For $a \sim 0.8$ we see one $H_0$ component exists; for $a \sim 0.5$ we see that one $H_0$ component takes birth along with an $H_1$ component; for $a \sim 0.4$ we see that the two $H_0$ components merge into one to cause the death of one of them. Finally for $a \sim 0.2$ the loop in the image fills up causing death of the $H_1$ component.

\subsubsection{Detection of Bifurcation using Homological Bifurcation Plots}

\begin{figure}[!htbp]
    \centering
    \includegraphics[width = 0.65\linewidth]{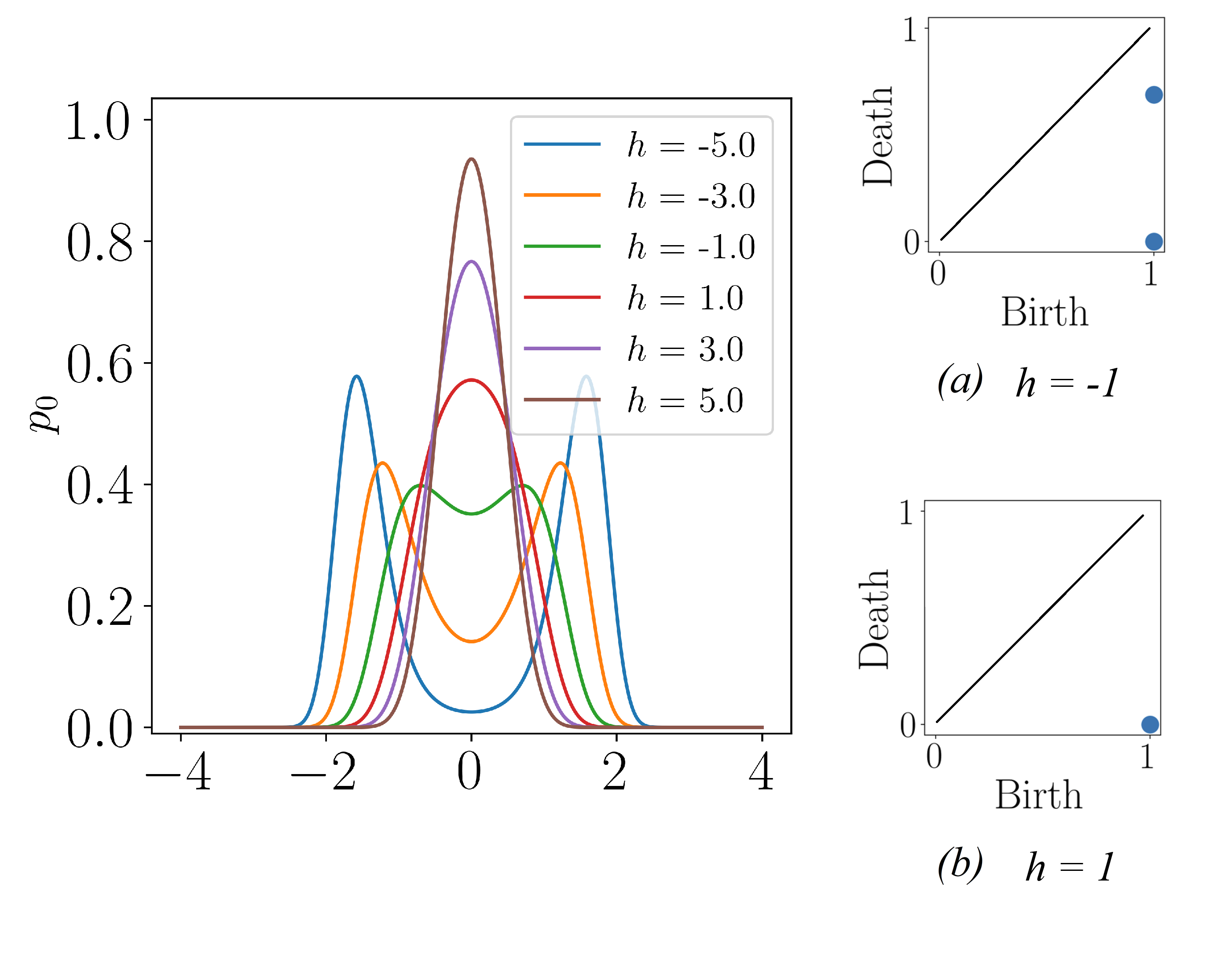}
    \includegraphics[width = 0.4\linewidth]{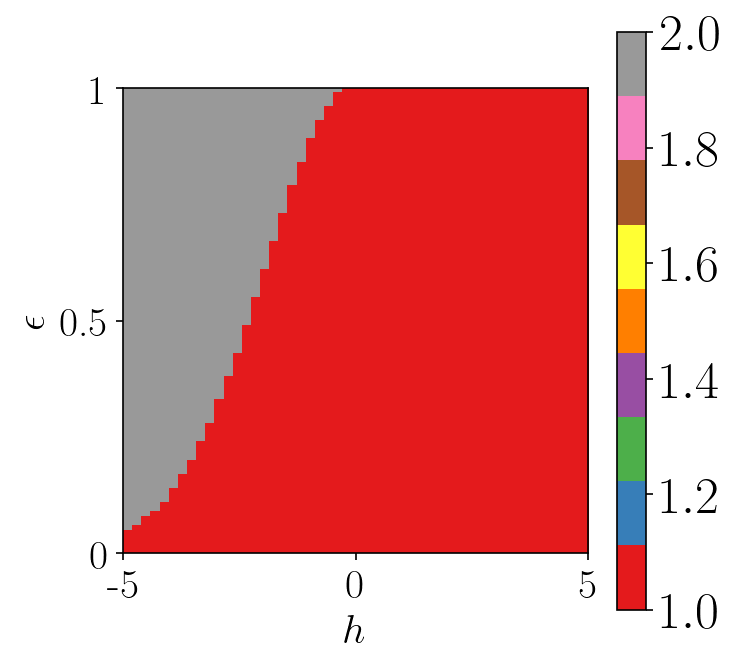}
    \includegraphics[width = 0.35\linewidth]{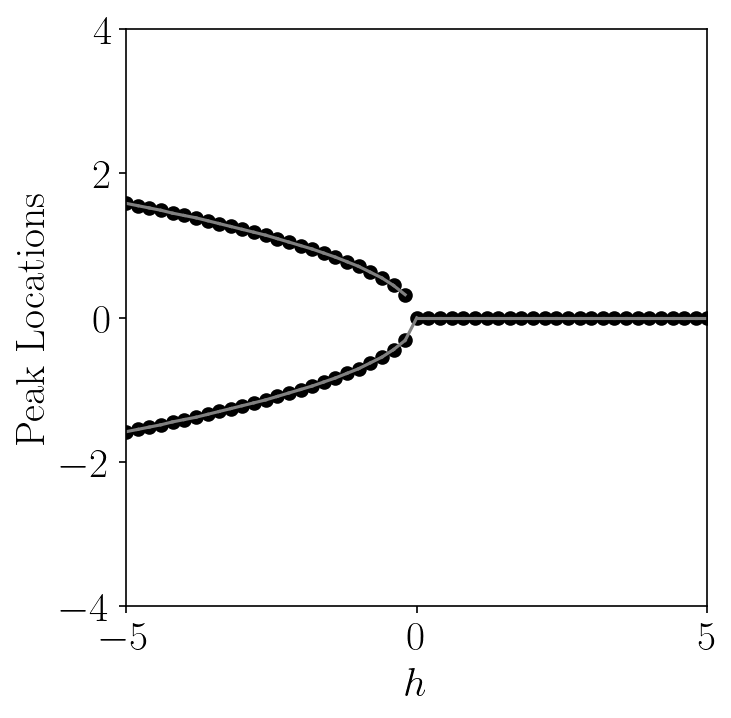}
    \caption{PDFs corresponding to various bifurcation parameter values in the function $p_x$. (a) The bistable PDF of $h=-1$ has two $H_0$ components in the persistence diagram while (b) the monostable PDF of $h=1$ has one $H_0$ component. }
    \label{fig:CubPerBif}
\end{figure}

A correspondence exists between the number of peaks in a PDF and the points in the cubical persistence diagram~\cite{Tanweer2024}, which allows for the detection of a phenomenological bifurcation by noting changes in colours in a homological bifurcation plot, see Fig.~\ref{fig:CubPerBif} for an example on a 1D PDF $p_x = C\exp{(-0.5*(x^4 + hx^2))}$ with $h \in [-5, 5]$. The homological bifurcation plot shows the change in consecutive columns at $h=0$ which is the bifurcation point, and the third plot tracks the $x$-location of the peaks against the bifurcation parameter.

\section{Results and Discussion}
\label{sec:R&D}

We analyzed the SIS model with parameters for two important diseases---gonorrhea and tuberculosis (TB)---and the SIR model with parameters for the common flu, in search of a bifurcation in the stationary PDF of $i(t)$ for SIS, and of  $r(t)$ for SIR, using three types of noise---white, Ornstein Uhlenbeck (OU) and logarithmic OU. See Table~\ref{tab:params} for the parameters used for the diseases.

\begin{table}[!htbp]
\centering
\begin{tabular}{||c c c c c||} 
 \hline
 $ $ & $R_0$ & $\gamma$ & $a$ & $\tau$\\ [0.5ex] 
 \hline\hline
 Gonorrhea & 1.4 & 1/month & 0.5 & -\\
 \hline
 Tuberculosis & 1.78 & 1/(6 month) & 0.5 & -\\ \hline
 Common Flu & 1.28 & 1/(7 day) & - & 1 day \& 7 day\\[1ex]
 \hline
\end{tabular}
\caption{Model parameters for the diseases of gonorrhea~\cite{Hethcote1984}, tuberculosis~\cite{Zhao2017} and common flu~\cite{Biggerstaff2014}. For the SIS diseases, i.e. gonorrhea and tuberculosis, we choose the relative correlation time $a=0.5$. For common flu which is a SIR disease, we choose as correlation time one day and one week motivated by the daily and weekly patterns in human activity~\cite{Kivel2012}.}
\label{tab:params}
\end{table}

\subsection{SIS Model}

In an SIS model, the stationary distribution of the infection population fraction $i(t)$ is a measure of the disease severity. A peak at non-zero location of the peak indicates a concentration of probability mass away from zero, making thus the endemic more probable, while a peak at zero indicates that disease eradication is more likely. The model parameters $R_0$ and $\gamma$ for gonorrhea and TB are given in Table~\ref{tab:params}; since also the timescale of an SIS model is easily determined (Sec.~\ref{subsubsec:comparison_OU_logOU}), we choose half of the SIS timescale as the correlation time of the OU and logOU noises. 

\begin{figure}[!htbp]
\centering
\includegraphics[width = 0.3\linewidth]{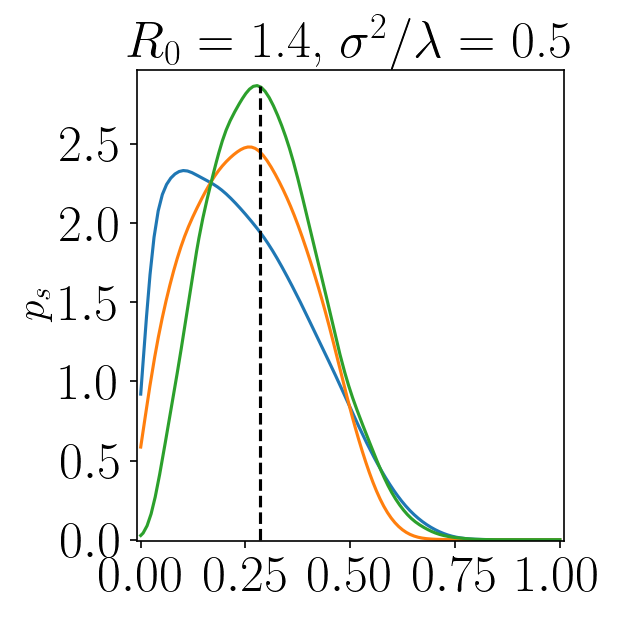}
\includegraphics[width = 0.3\linewidth]{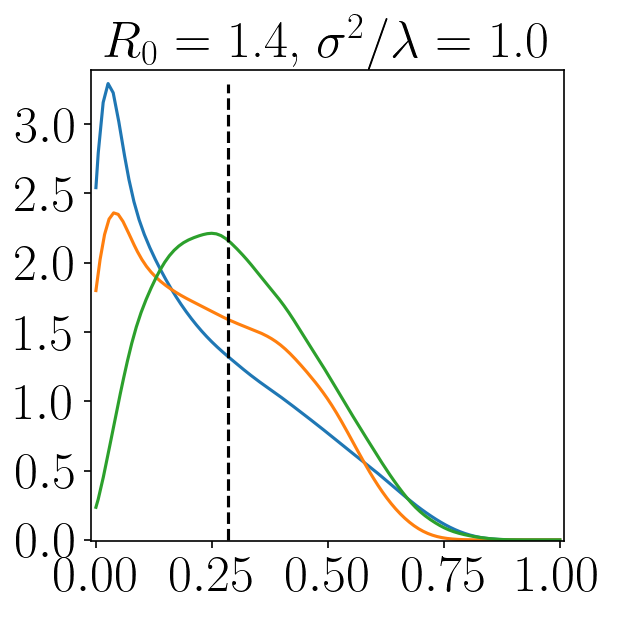}
\includegraphics[width = 0.3\linewidth]{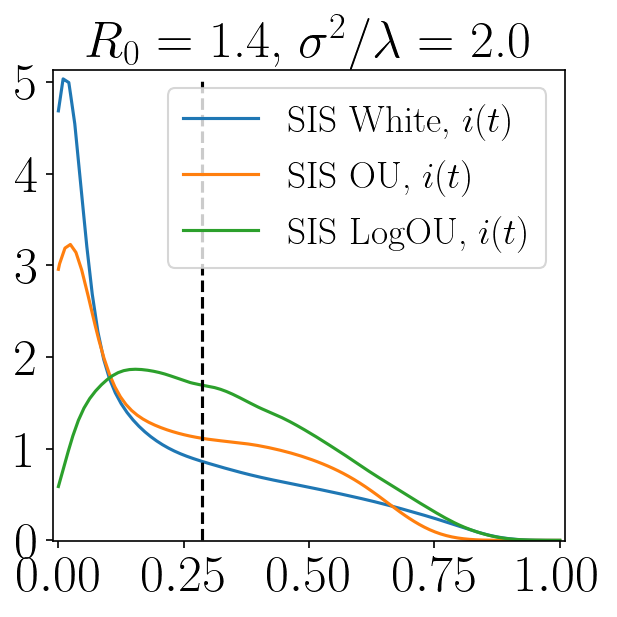}
\includegraphics[width = 0.3\linewidth]{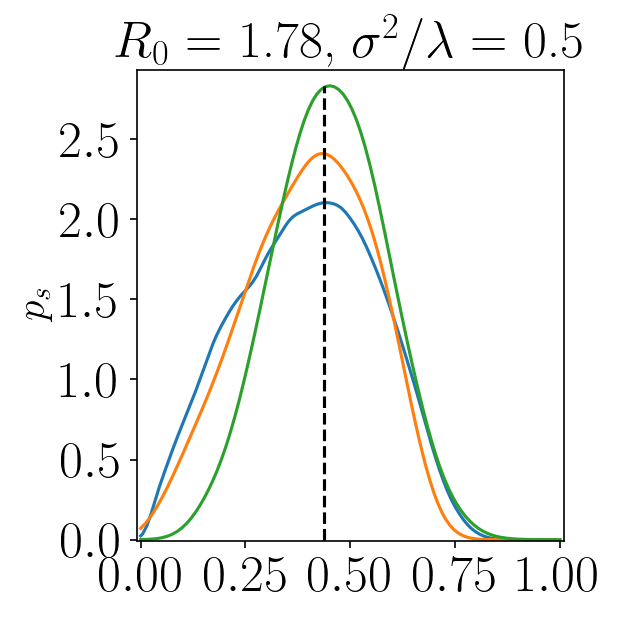}
\includegraphics[width = 0.3\linewidth]{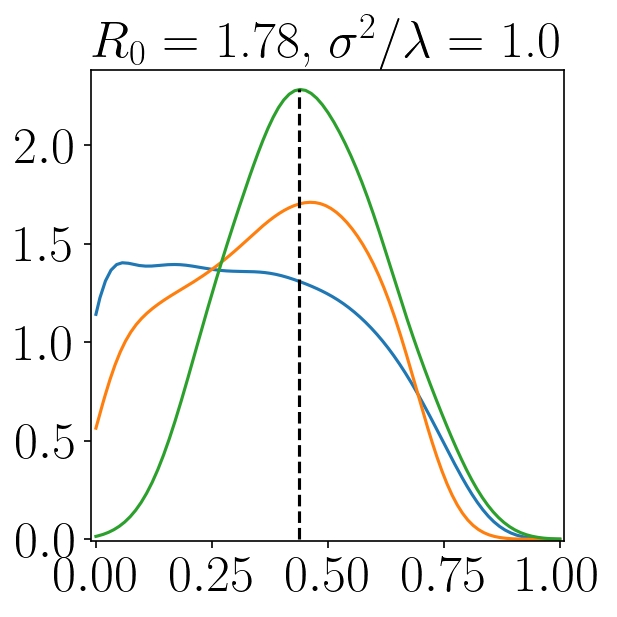}
\includegraphics[width = 0.3\linewidth]{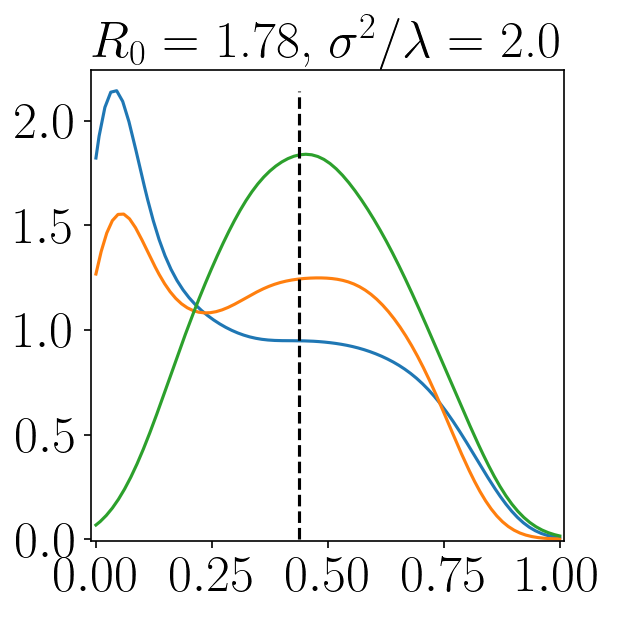}
\caption{KDEs for $i(t)$ for diseases of gonorrhea ($R_0 = 1.4$) and TB ($R_0 = 1.78$) with different relative noise intensities for all three noise types.}
\label{fig:overview_SIS}
\end{figure}

Figure~\ref{fig:overview_SIS} explores the KDEs of the infected fraction for low, moderate and high noise intensities. For both diseases (different $R_0$), the KDE is plotted for all three noise types. The deterministic fixed point is represented as a dotted line. 
For both diseases as noise intensity increases, white and OU noise simulations show pronounced peaks at 0, indicating an increasing probability of disease eradication. In contrast, for logarithmic OU noise, the distribution never has a peak at zero. For gonorrhea, the peak moves much closer to 0 but is still located at a non-zero value while for TB, the peak is strongly fixed at the deterministic fixed point. The spread in the distribution increases with the noise intensity for both diseases, as expected.

\begin{figure}[!htbp]
\centering

\includegraphics[width = 0.4\linewidth]{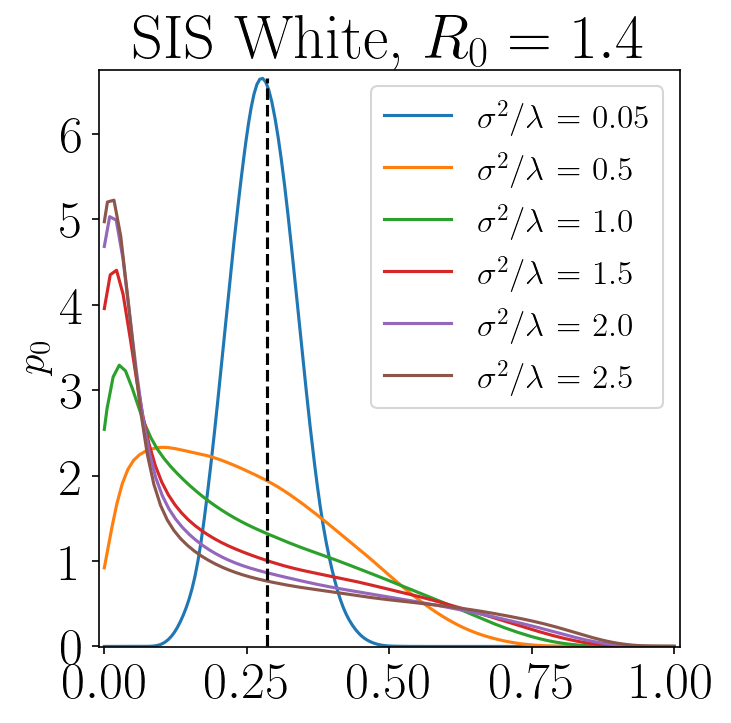}
\includegraphics[width = 0.4\linewidth]{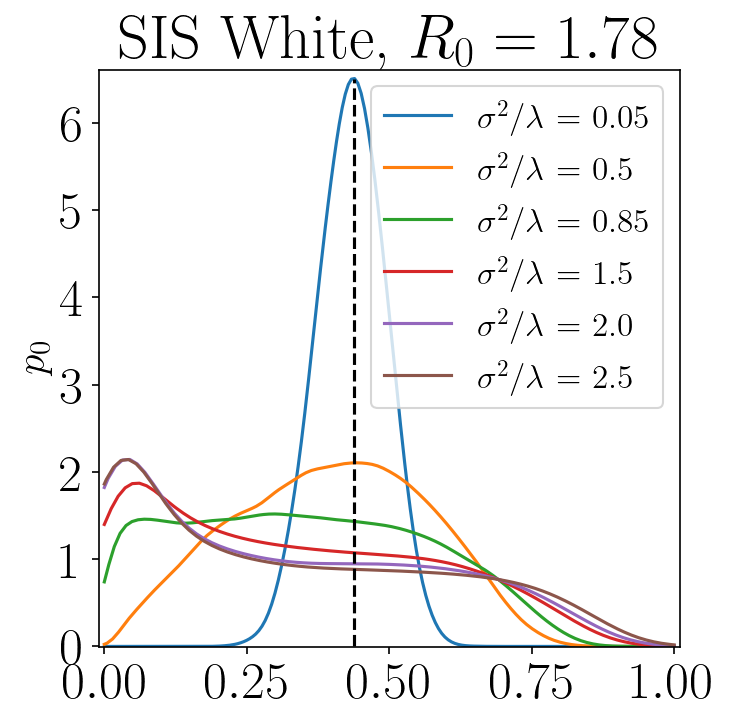}
\\
\includegraphics[width = 0.4\linewidth]{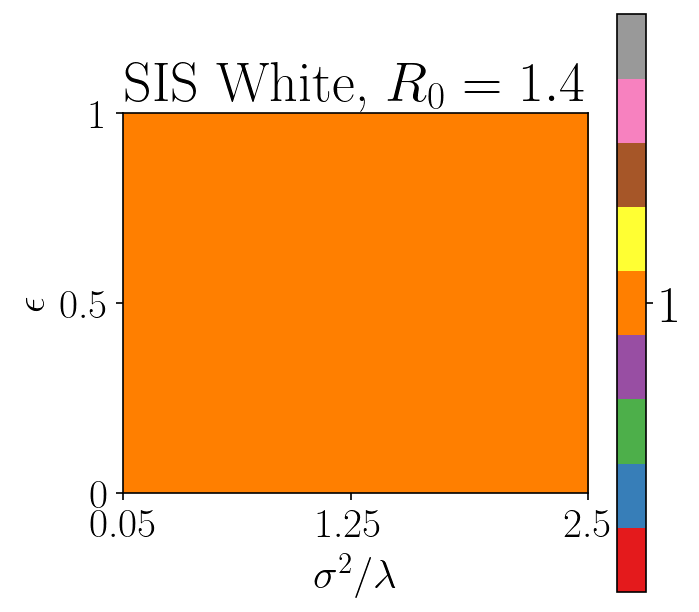}
\includegraphics[width = 0.4\linewidth]{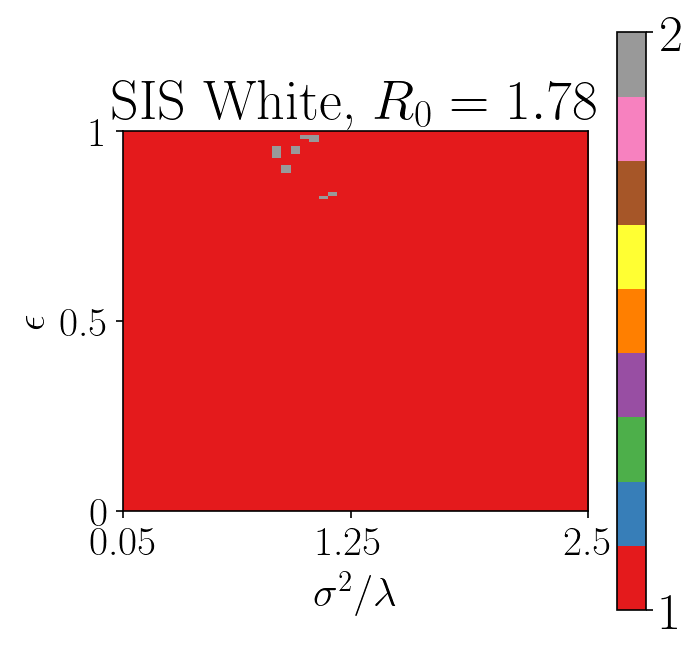}
\\
\includegraphics[width = 0.4\linewidth]{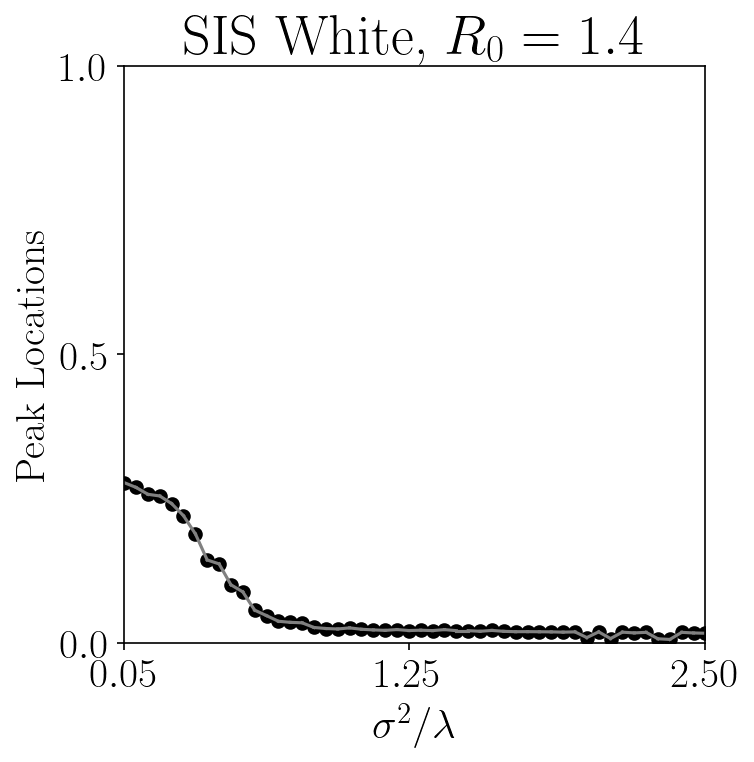}
\includegraphics[width = 0.4\linewidth]{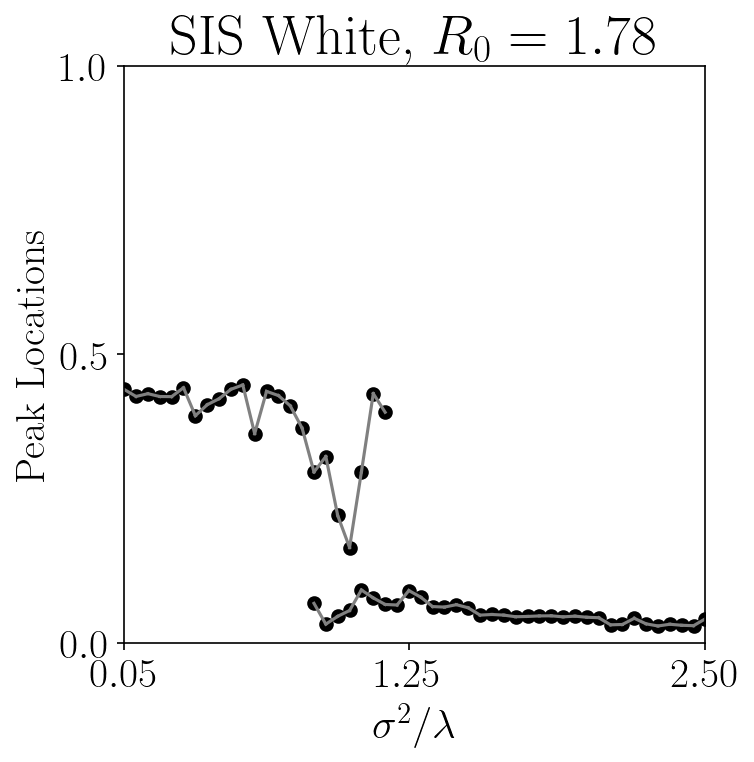}
    
\caption{Steady-state KDEs for SIS model over a range of relative noise intensities under white noise for gonorrhea and TB, with corresponding bifurcation and peak-tracking plots.}

\label{fig:intensities_SIS_White}
\end{figure}

Figure~\ref{fig:intensities_SIS_White} investigates the effect of increasing relative noise intensity on the KDE of the infected fraction under white noise only, for both diseases, alongwith the corresponding homological bifurcation plots---which track the number of peaks in the distribution---and plots tracking the location of all peaks. For gonorrhea, the deterministic fixed point is approximately 0.27, but under all noise intensities except very low, the KDEs exhibit a single dominant peak at 0.
As the noise intensity increases, the peak at 0 becomes sharper, reflecting a higher probability of disease eradication. 
The homological bifurcation plot confirms that the system consistently maintains a single peak across all noise intensities, and the location-tracking plot shows the peak slowly shifting to a location of 0 with increasing noise. For TB, the deterministic fixed point lies at about 0.45. For comparatively low relative noise intensites, the KDE shows a single peak located between 0 and the deterministic fixed point. As the noise intensity increases, the peak shifts closer to 0, and the probability mass at 0 grows. This indicates that higher noise levels suppress the infected fraction, increasing the likelihood of eradication. The homological bifurcation plot indicates a single peak across all noise intensities except for a small range of intensities between 0.8 and 1.2 where two peaks exist---corroborating the KDE observations. The location-tracking plot also shows two peaks in the same noise interval with one peak for all other instances. These plots highlight how noise intensity influences the stability and behavior of the system under white noise. 

\begin{figure}[!htbp]

\centering
\includegraphics[width = 0.4\linewidth]{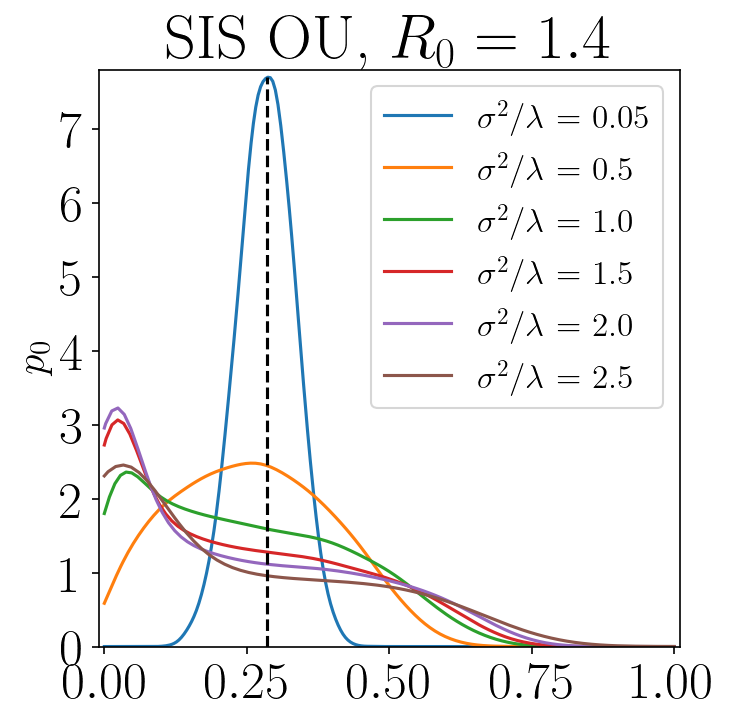}
\includegraphics[width = 0.4\linewidth]{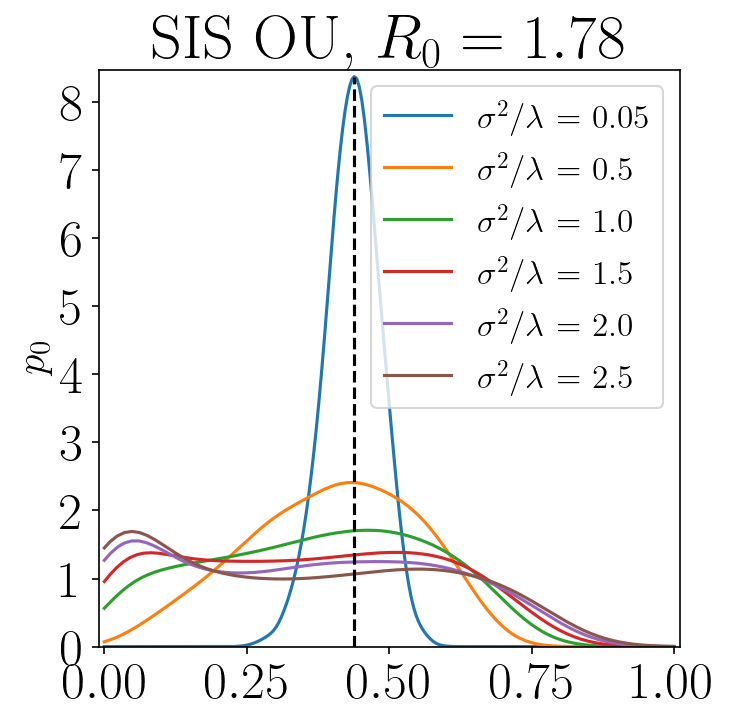}
\\
\includegraphics[width = 0.4\linewidth]{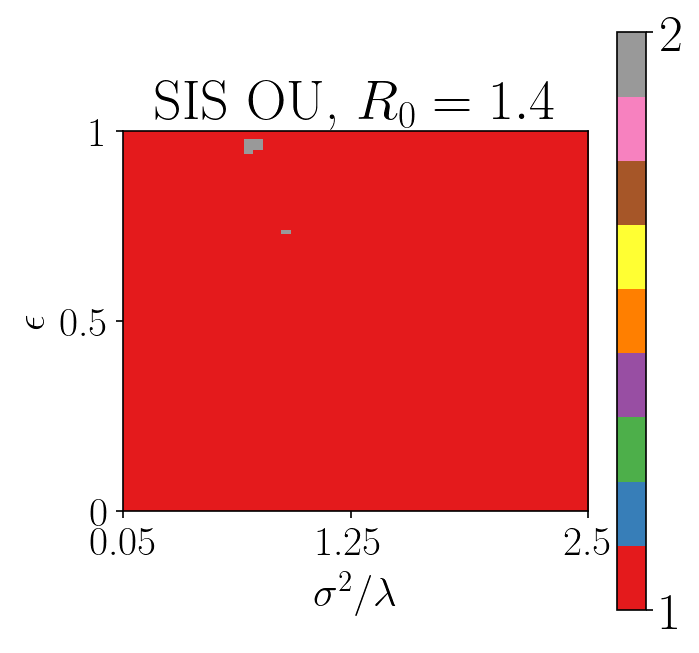}
\includegraphics[width = 0.4\linewidth]{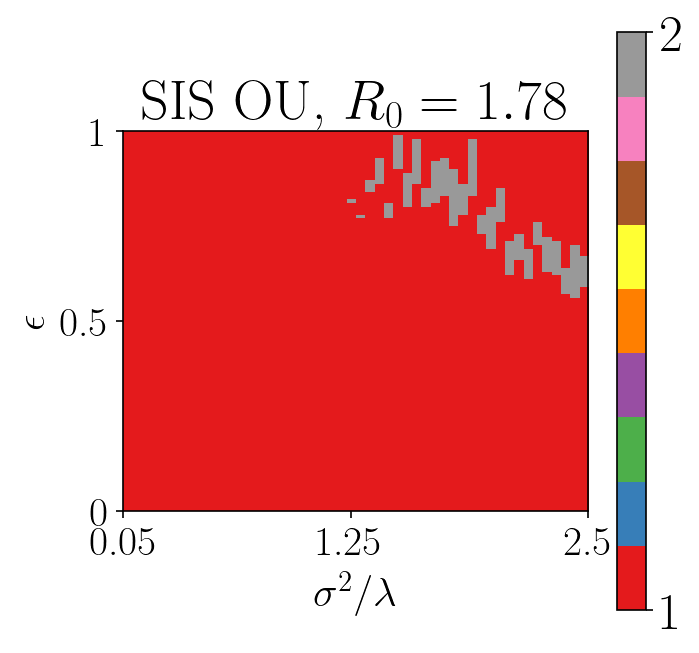}
\\
\includegraphics[width = 0.4\linewidth]{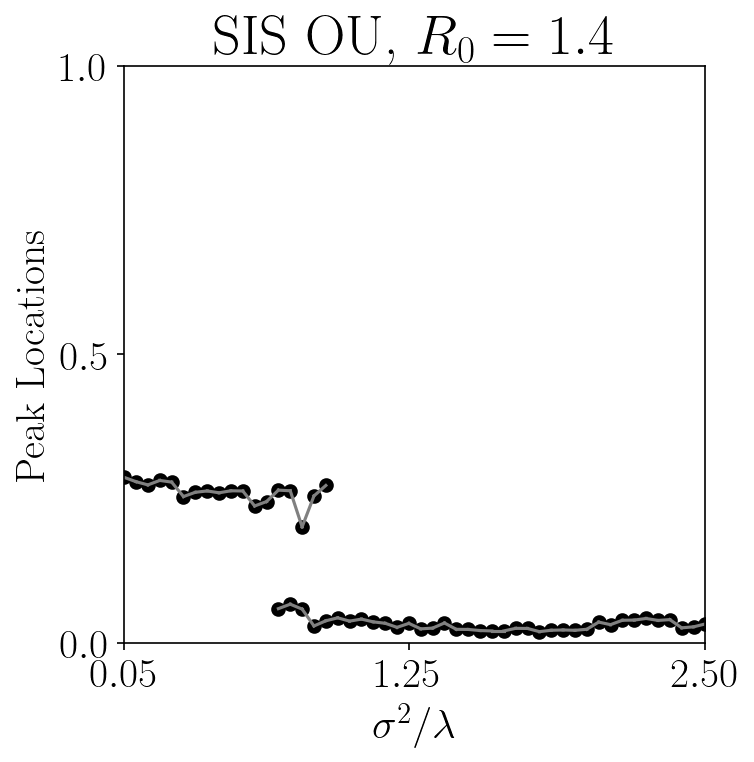}
\includegraphics[width = 0.4\linewidth]{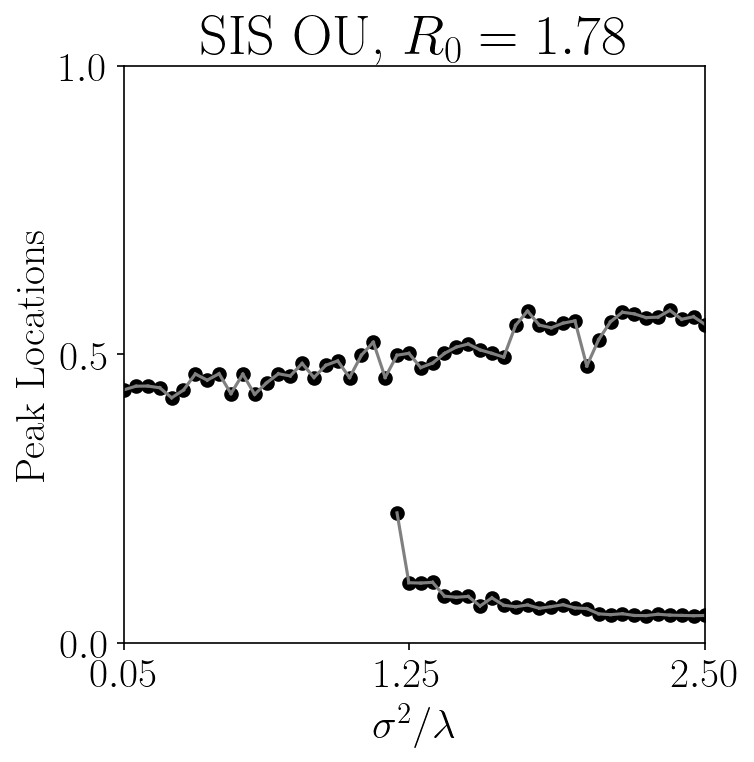}

\caption{Steady-state KDEs for SIS model over a range of relative noise intensities under OU noise for gonorrhea and TB, with corresponding bifurcation and peak-tracking plots.}

\label{fig:intensities_SIS_OU}
\end{figure}

Figure~\ref{fig:intensities_SIS_OU} shows the same cases as Fig.~\ref{fig:intensities_SIS_White} but with OU excitation. The general trend and behavior with OU noise share many similarities with white noise dynamics; however, the emergence of the peak at zero happens for higher noise intensities for OU noise. For $R_0$ of 1.4, the infected fraction displays a single dominant peak between 0 and the deterministic fixed point at low noise. As the noise intensity grows, another peak develops close to 0. For higher noise, the peak at a non-zero value dies leaving only the peak at 0. For $R_0$ of 1.78, the KDE has a single peak which slowly shifts to a non-zero value above the deterministic fixed point, while a new peak is induced at location of 0 as the noise intensity increases beyond 1.0. The homological bifurcation plot show this second peak popping up as the gray region while the location-tracking plot depict them with a new line around $\sigma^2 / \lambda \geq 1.0$.

\begin{figure}[!htbp]

\centering
\includegraphics[width = 0.4\linewidth]{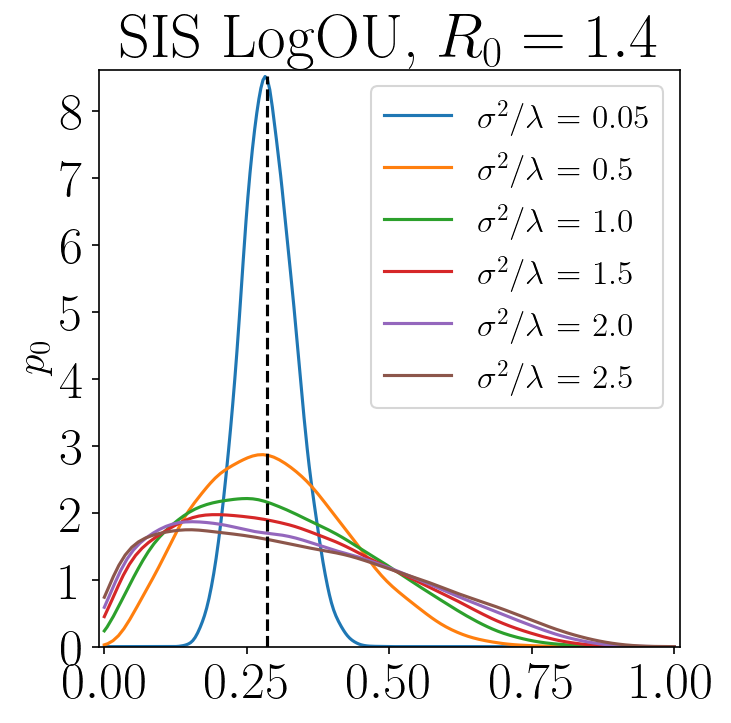}
\includegraphics[width = 0.4\linewidth]{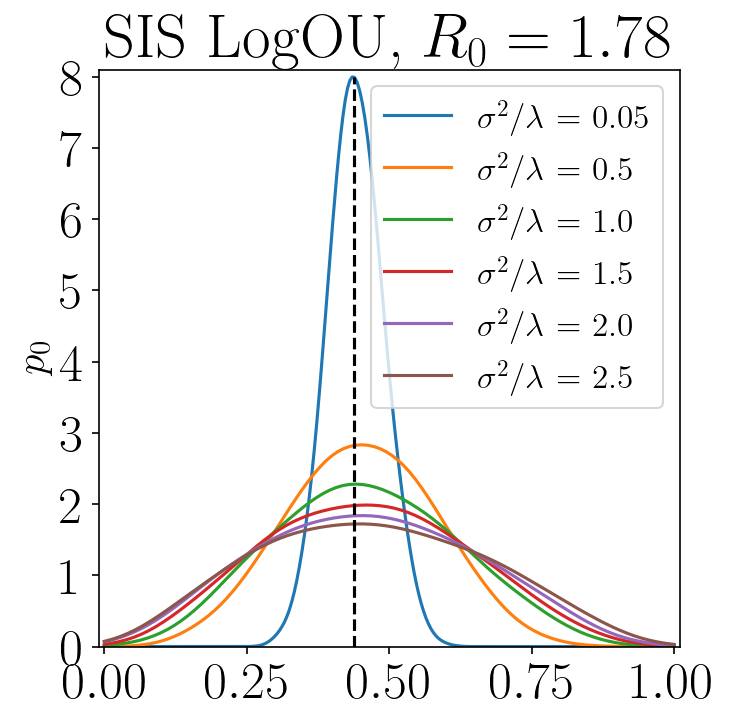}
\\
\includegraphics[width = 0.4\linewidth]{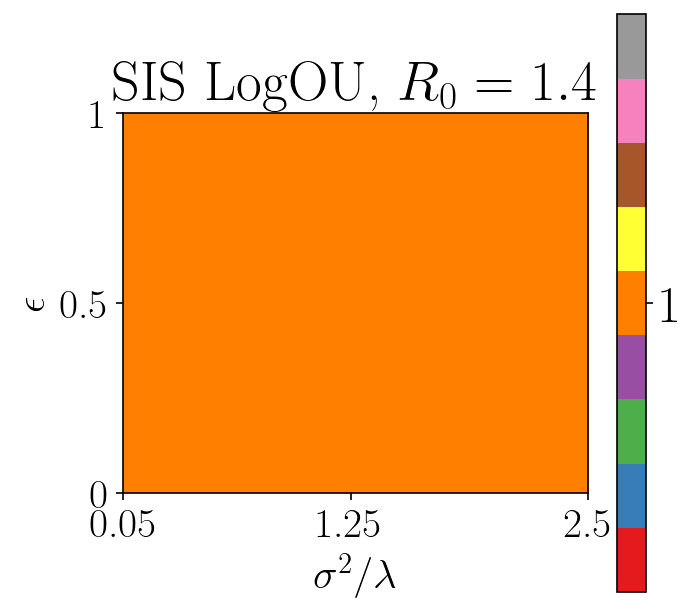}
\includegraphics[width = 0.4\linewidth]{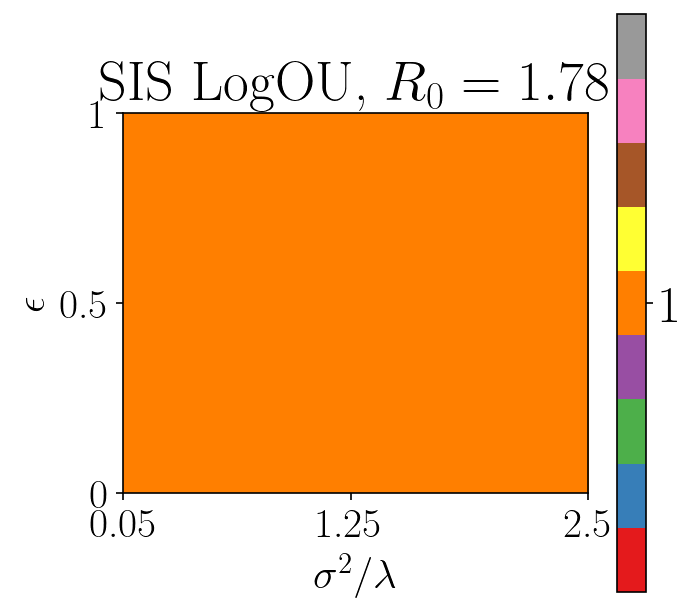}
\\
\includegraphics[width = 0.4\linewidth]{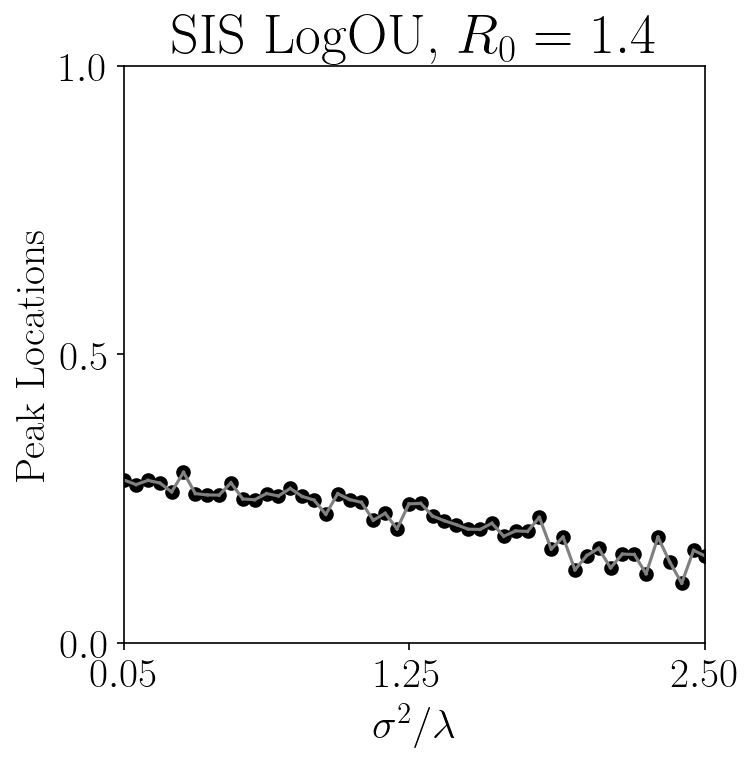}
\includegraphics[width = 0.4\linewidth]{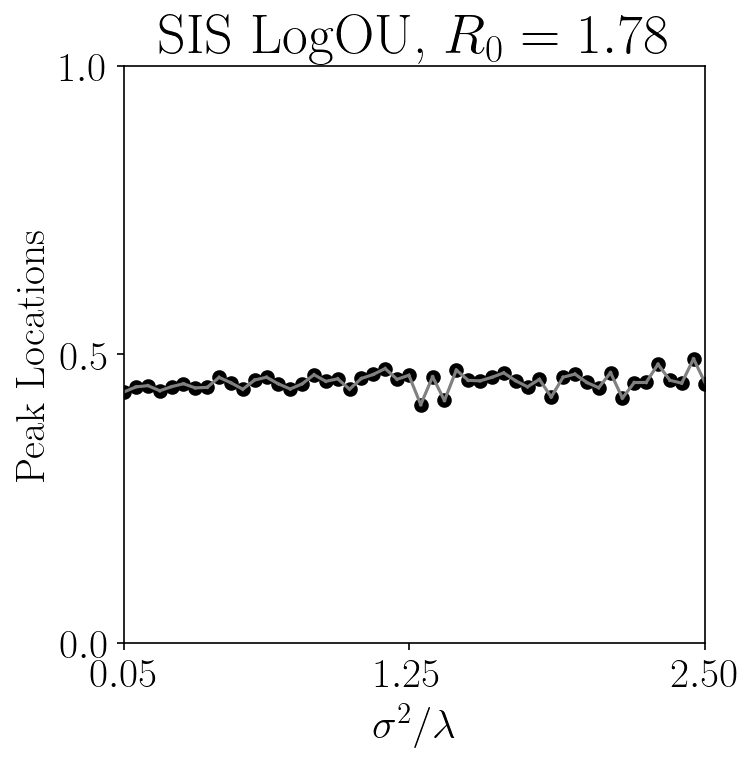}

\caption{Steady-state KDEs for SIS model over a range of relative noise intensities under logarithmic OU noise for gonorrhea and TB, with corresponding bifurcation and peak-tracking plots.}

\label{fig:intensities_SIS_LogOU}
\end{figure}

The behavior of the SIS model under logarithmic OU noise with varying relative noise intensities is illustrated in Figure~\ref{fig:intensities_SIS_LogOU} for gonorrhea and TB. Unlike the cases with white and OU noise, no secondary peak is observed in the KDEs regardless of the noise intensity. For $R_0$ of 1.4, the KDEs show a single peak slightly to the left of the deterministic fixed point. As the noise intensity increases, the endemic peak drifts towards zero, indicating a higher likelihood of disease eradication. For $R_0$ of 1.78, the KDEs also show a single peak but very slightly to the right of the deterministic fixed point. As the noise intensity increases, the variance in the KDEs increases, as expected. In both cases, the homological bifurcation plots show no change---proving that no bifurcation occurs since no second peak emerges. These results suggest that under logarithmic OU noise, due to its all-positive nature, the system shows greater resilience to noise-induced transitions compared to white and OU noise, and the noise strongly stabilizes the system near its deterministic behavior, even at high intensities.

\subsection{SIR Model}

SIR models compartmentalize the population into Susceptible ($s(t)$), Infected ($i(t)$) and Removed ($r(t)$) populations. In these, the KDE of $r(t)$ can be used to understand the disease severity. 

\begin{figure}[!htbp]
\centering
\includegraphics[width = 0.3\linewidth]{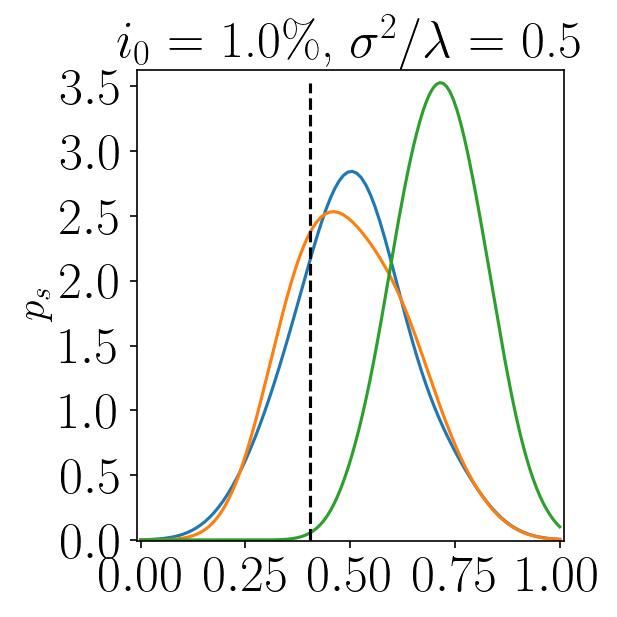}
\includegraphics[width = 0.3\linewidth]{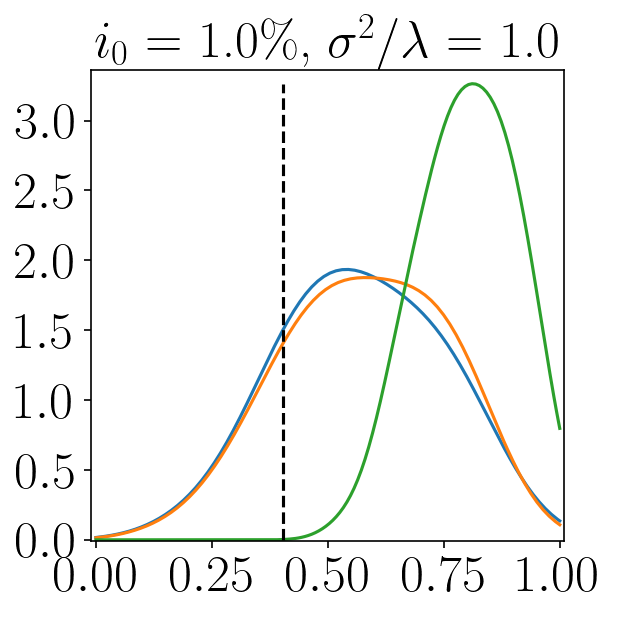}
\includegraphics[width = 0.3\linewidth]{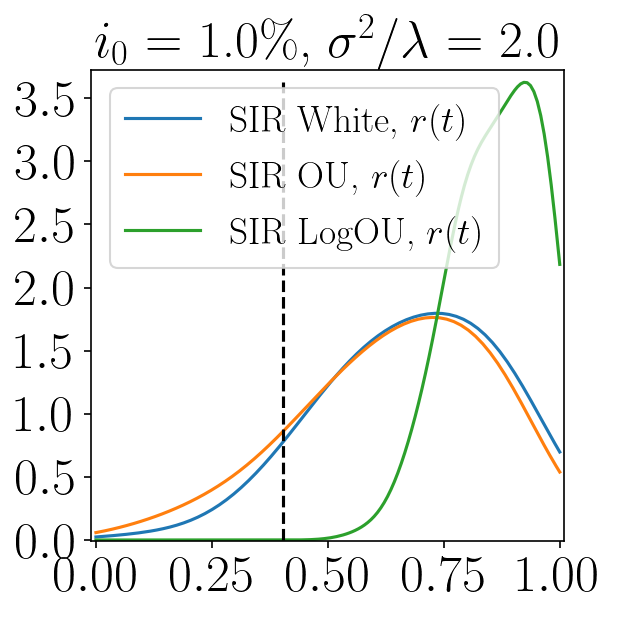}

\includegraphics[width = 0.3\linewidth]{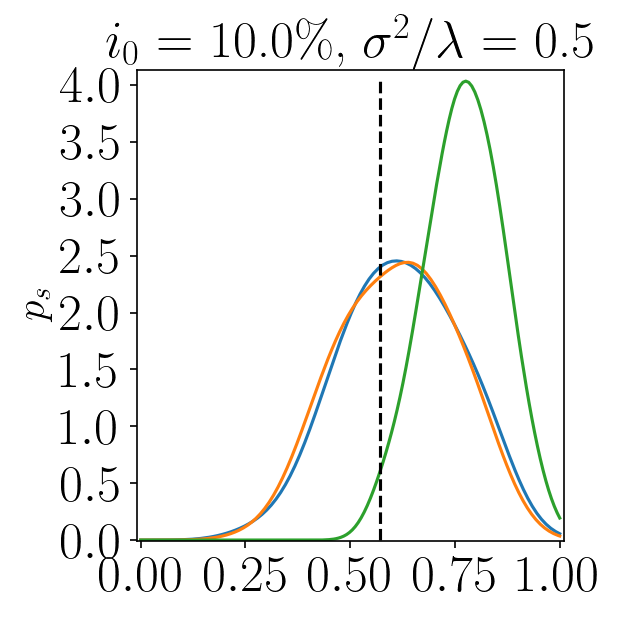}
\includegraphics[width = 0.3\linewidth]{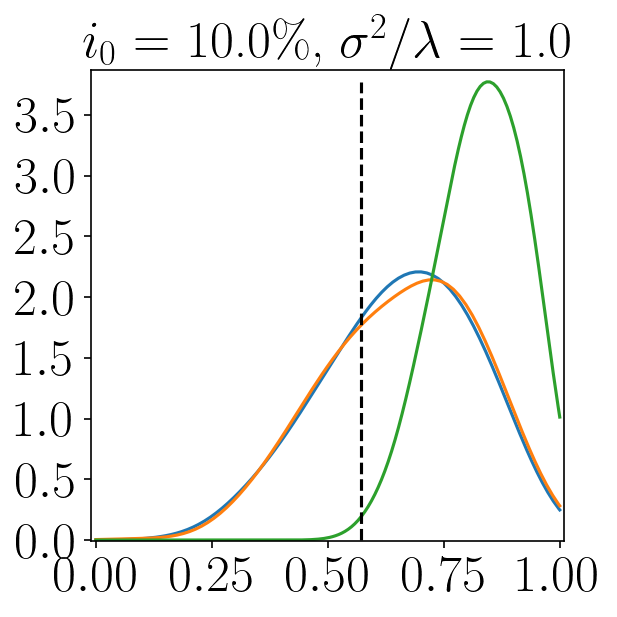}
\includegraphics[width = 0.3\linewidth]{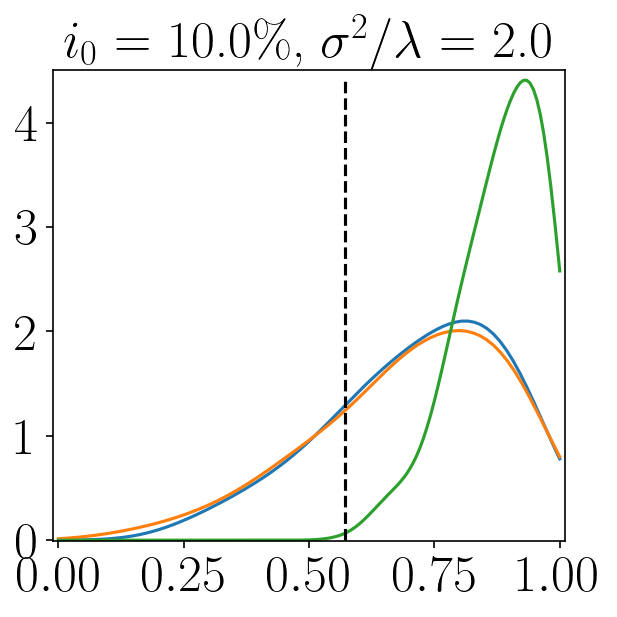}

\caption{Steady-state KDEs for flu ($R_0 = 1.28$) with $\tau = 1$ day.}
\label{fig:overview_SIR_1}

\end{figure}

Figure~\ref{fig:overview_SIR_1} explores the KDEs of the removed fraction under SIR dynamics for common flu with correlation time $\tau = 1$ day for low, moderate and high relative noise intensities. The top row is simulated for a low initial infected population of $1\%$ while the bottom row is for a high initial infected population of $10\%$. For both cases, the KDE is plotted for all three noise types and the deterministic fixed point is represented as a dotted line. For these parameter values, all noise types exhibit only a monostable response. However, logarithmic OU noise KDEs have their peaks shifted significantly to the right of the deterministic fixed point. In contrast, for white and OU, the distribution's peak is only slightly to the right of the deterministic fixed point. In all cases, either no or very little probability mass exists in the vicinity of 0. 

\begin{figure}[!htbp]
\centering
\includegraphics[width = 0.8\linewidth]{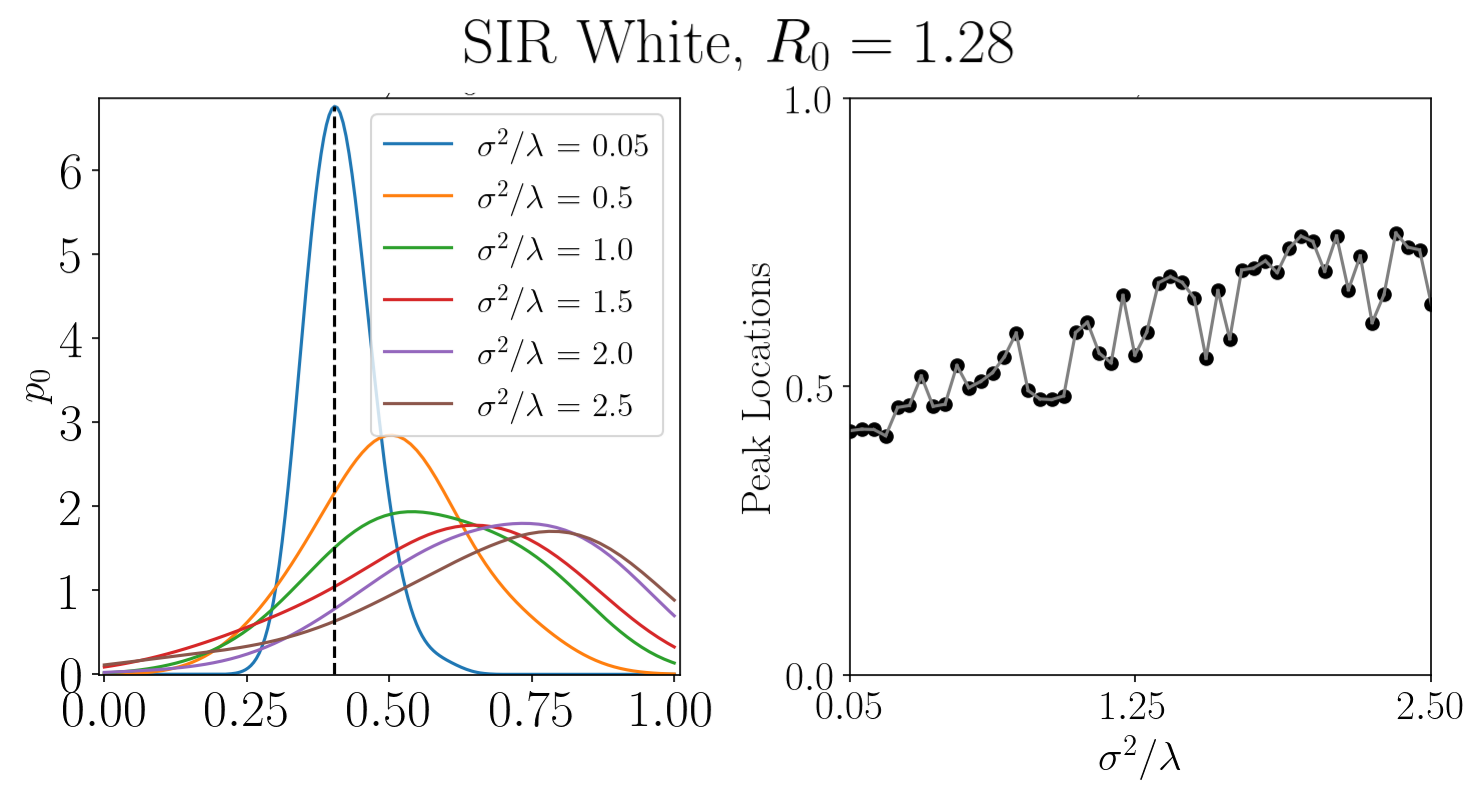}
\caption{Steady-state KDEs for SIR model over a range of relative noise intensities under white noise for flu with initial infected populations of $1\%$, with corresponding peak-tracking plot.}

\label{fig:intensities_SIR_White_1}
\end{figure}

Figure~\ref{fig:intensities_SIR_White_1} illustrates the behavior of the SIR model under white noise. In all cases, the KDEs remain monostable and ahead of the deterministic fixed point except for a very low noise intensity. In all cases, the peak drifts further away from the deterministic fixed point as the noise intensity increases. There is effectively no probability mass at the value 0 in all cases. The bifurcation plots agree with the KDEs showing a single peak and no bifurcation, while the peak location plots accurately depict the peaks' drift towards 1 with increasing noise intensity.

\begin{figure}[!htbp]
    \centering
\includegraphics[width = 0.8\linewidth]{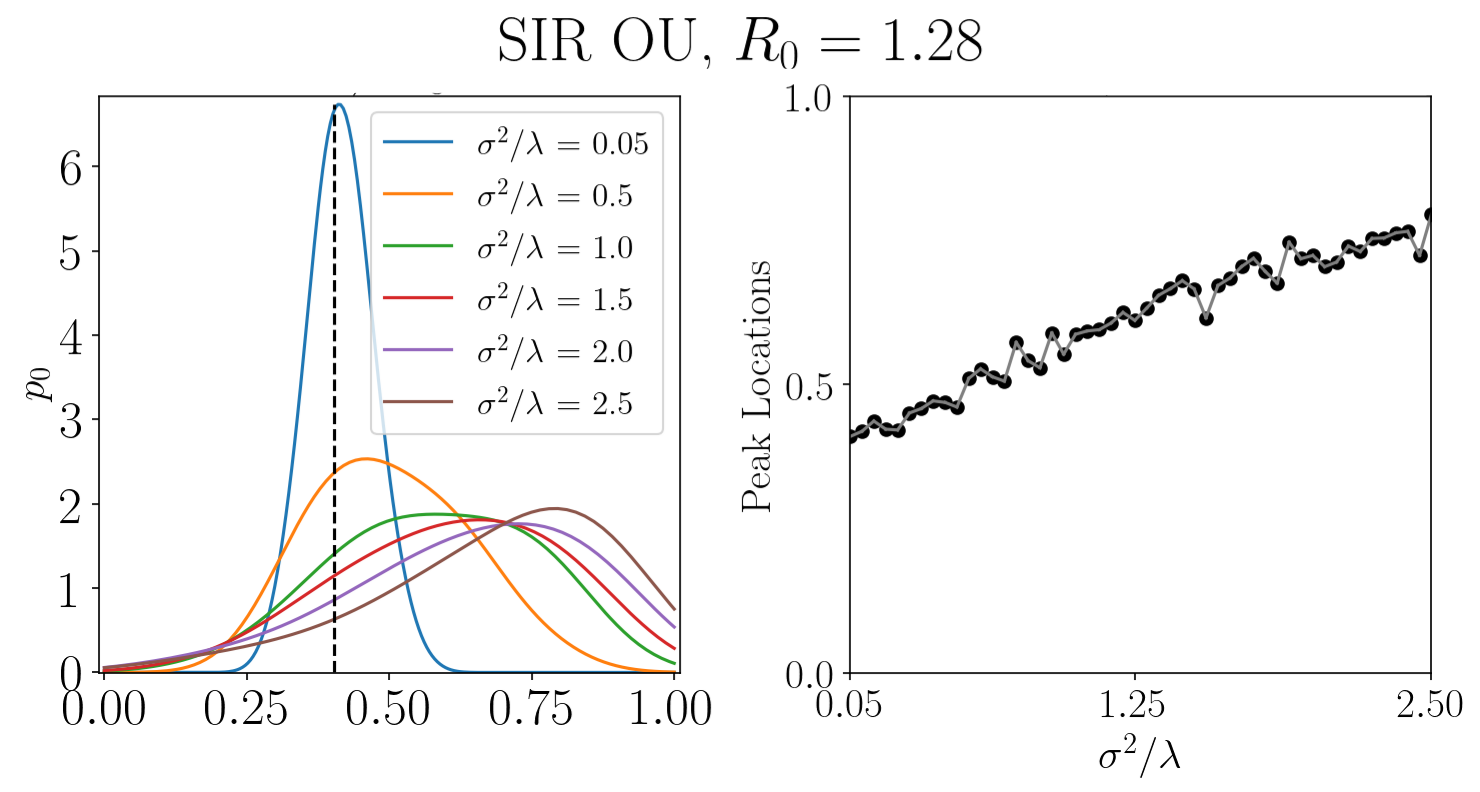}

\caption{Steady-state KDEs for SIR model over a range of relative noise intensities under OU noise for flu with $\tau = 1$ and initial infected populations of $1\%$, with corresponding peak-tracking plot.}
    \label{fig:intensities_SIR_OU_1}
\end{figure}

Figure~\ref{fig:intensities_SIR_OU_1} illustrates the behavior of the SIR model under OU noise for $\tau = 1$ day. In all cases, the KDEs remain monostable. The peak can be seen to drift away from the deterministic fixed point as the noise intensity is increased with an increasing variance in the distribution. There is virtually no probability mass at the value of 0 in all cases.

\begin{figure}[!htbp]
    \centering
\includegraphics[width = 0.8\linewidth]{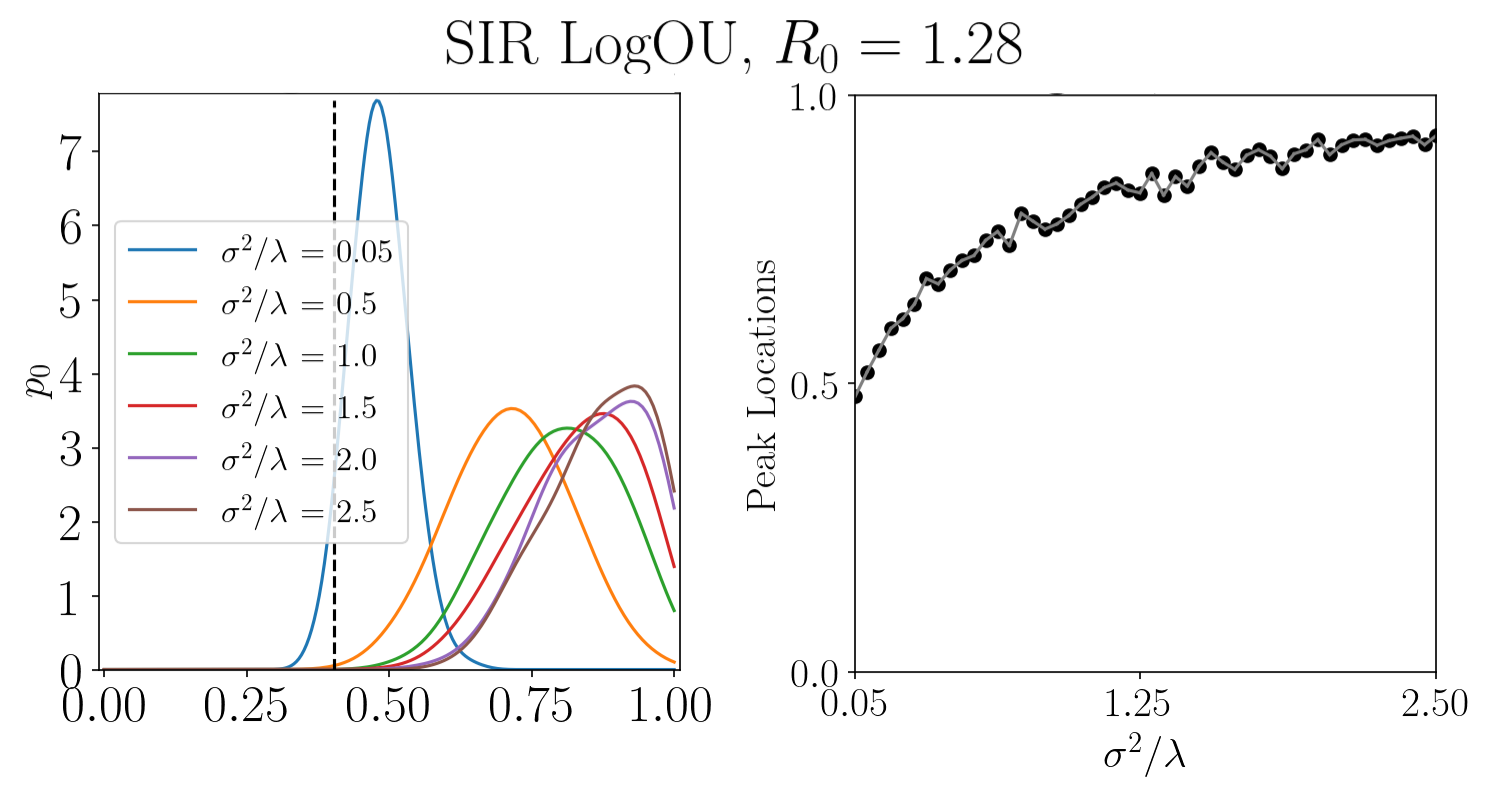}

\caption{Steady-state KDEs for SIR model over a range of relative noise intensities under LogOU noise for flu with $\tau = 1$ and initial infected populations of $1\%$, with corresponding peak-tracking plot.}
\label{fig:intensities_SIR_LogOU_1}

\end{figure}

Finally, the behavior of the SIR model under logarithmic OU noise is shown in Fig.~\ref{fig:intensities_SIR_LogOU_1}. The dynamics under logarithmic OU follow a similar trend as the OU noise. However, the drift in the peaks' location seems faster for logarithmic OU noise. Other observations remain the same---increase in variance with increase in noise and virtually zero probability mass at 0. 

A similar analysis for SIR model with $\tau = 7$ days is given in Appendix~\ref{sec:appB}. The overall dynamics and bifurcation behavior appear to be the same. 

\section{Conclusion}
\label{sec:Conc}

Stochastic compartmental models are important to understand the mechanisms for the eradication of diseases in a population. They are also important in predicting whether a disease will become endemic or not. In this work, we used compartmental models with stochastic contact rate. Three different noise types have been utilized: white, OU and logarithmic OU. We analyzed the system for various basic reproduction numbers and relative noise intensities, searching for noise-induced phenomenological bifurcations that would result in probability mass concentrating around zero and thus making disease eradication more likely. 

While white noise has been extensively used for stochastic perturbation of contact rate, and OU noise is able to incorporate correlations that arise from the patterns in human social activity, both noises have a main drawback. They are Gaussian and thus unbounded, which may result in the fluctuating contact rate attaining negative values, which is physically impermissible, especially in cases of high noise intensity. For this reason, the always positive logOU noise has been proposed instead. However, for the logOU noise, the analytical tools for solving compartmental models under Gaussian perturbations no longer apply. Thus, we resort to simulation-based homological bifurcation plots, which provide a robust framework for detecting and visualizing noise-induced transitions while also capturing the drift in peak locations. 

For SIS models, both white and OU noise show similar trends, with secondary peaks at zero emerging at high noise intensities. However, the onset of these peaks occurs earlier for white noise, indicating higher susceptibility to noise-induced bifurcations. In contrast, logarithmic OU noise demonstrates significant robustness, maintaining a unimodal structure across all cases investigated, with a peak at a non-zero point (endemic peak). For SIR models, similar trends were observed for all three noise types, with no peak at zero emerging, and slow drift in the peaks' location towards higher values. Compared to the white and OU noises, KDEs under logarithmic OU noise show less variance in the distributions.

From our results, we conclude that the positivity of logOU noise makes the predictions of compartmental models more robust. Thus, the emergence of the secondary peak at zero most likely originates from the probability of the contact rate perturbed by white and OU noise to attain negative values, which is increasing as the noise intensity increases. However, we observe the same trend in the drift of the endemic peak in all noises: In SIS models and for relatively low reproduction numbers $R_0$ as in the example of gonorrhea, increasing the noise intensity results in the endemic peak drifting toward zero for all three types of noise, making disease eradication more likely. In contrast, for high $R_0$ as in the example of TB, increasing the noise intensity causes the endemic peak to drift towards higher values, making the disease outbreak more severe. This is a novel result towards the qualitative understanding of the different effects of noise in disease spread, for diseases with different reproduction numbers.
\section{Acknowledgements}
This material is based on work supported by the Air Force Office of Scientific Research under award number FA9550-22-1-0007.

\newpage
\printbibliography
\newpage
\begin{appendices}
\section{Noise Properties for High Intensity}
\label{sec:appA}

\begin{figure}[!htbp]
\centering
\includegraphics[width=0.3\linewidth]{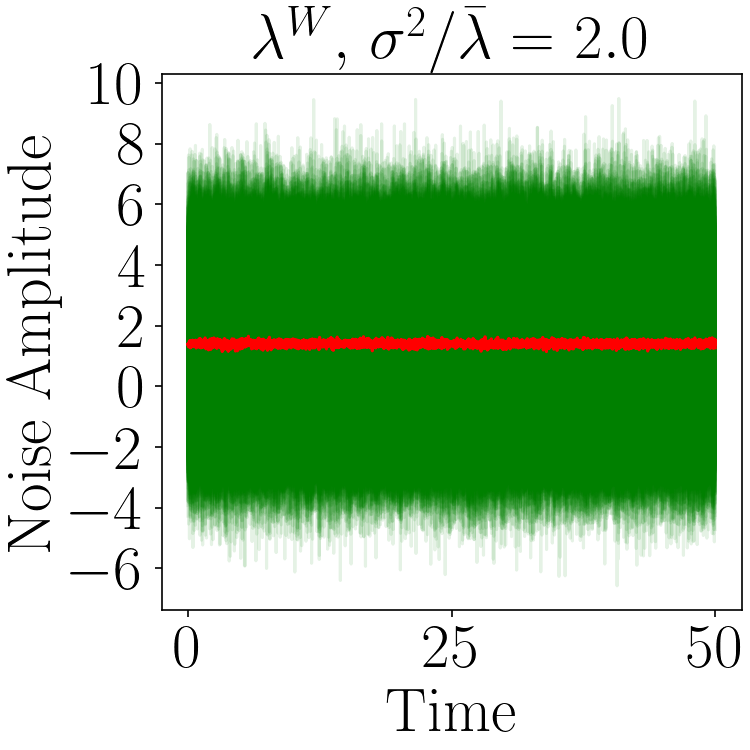}
\includegraphics[width=0.3\linewidth]{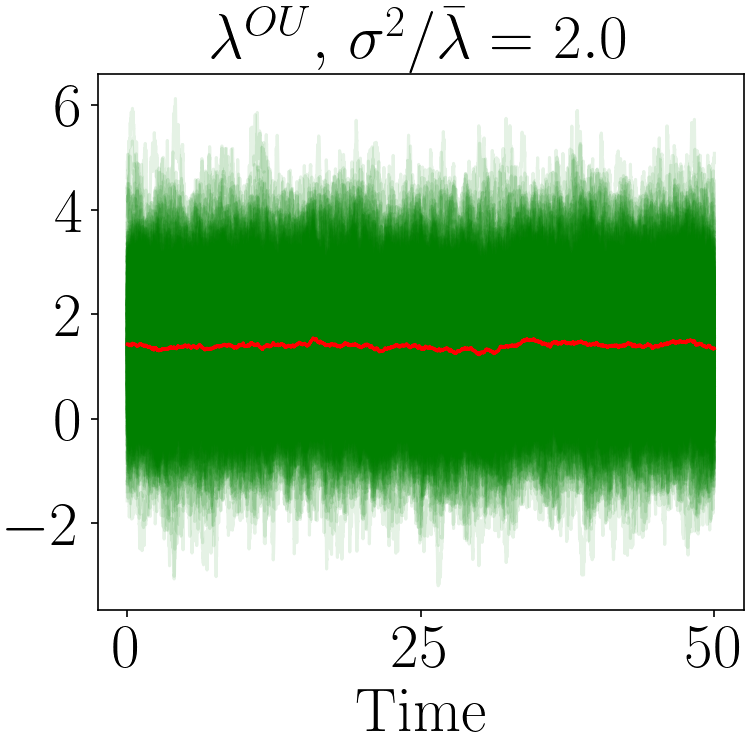}
\includegraphics[width=0.3\linewidth]{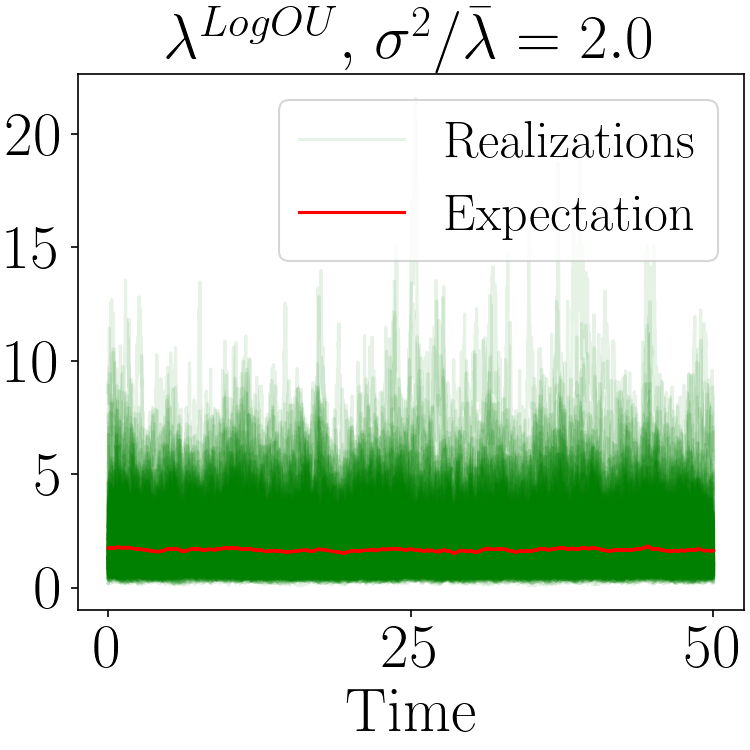}
\\
\includegraphics[width=0.3\linewidth]{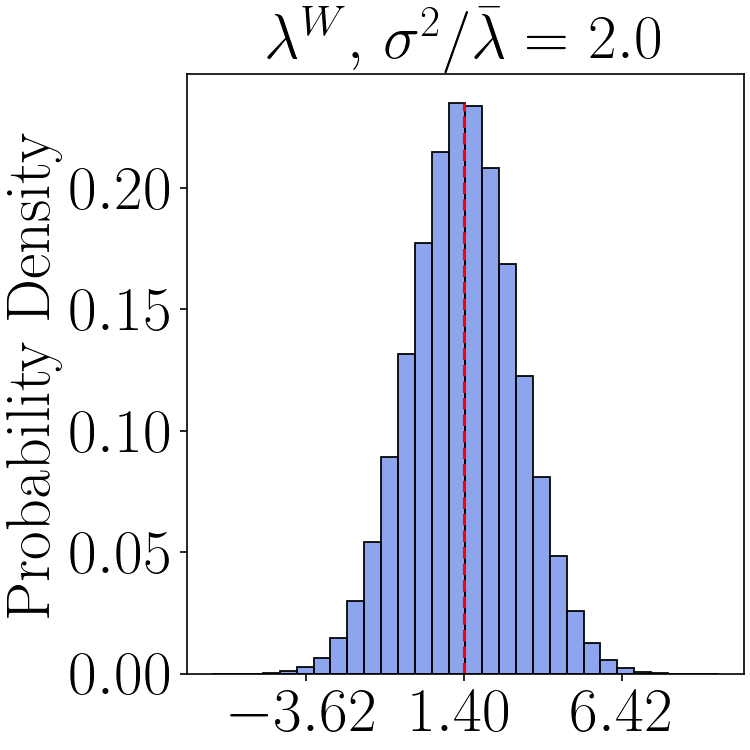}
\includegraphics[width=0.3\linewidth]{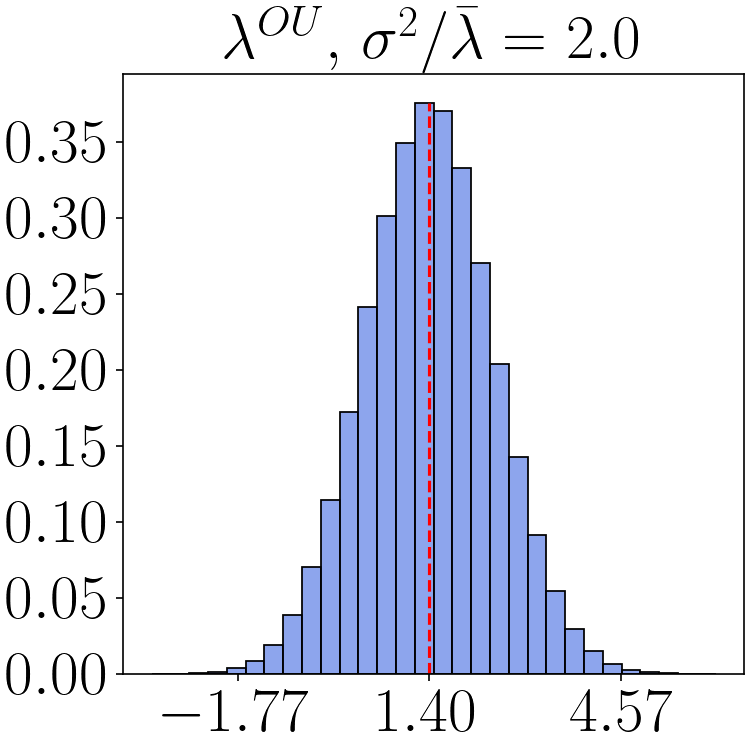}
\includegraphics[width=0.3\linewidth]{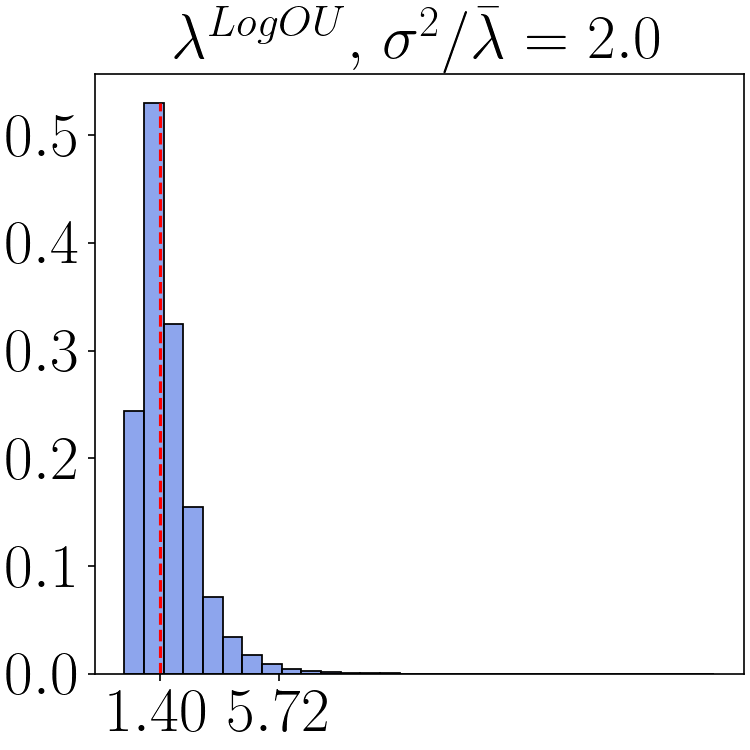}
\caption{Trajectories for contact rate $\lambda$ having $\bar{\lambda} = 1.4 \text{ per month}$ with white, OU and logarithmic OU noise perturbations, followed by the distributions for each---for a relative noise intensity of 2.0. Basic reproduction number $R_0 = 1.4$ (gonorrhea) was taken for these simulations.}
\end{figure}

\begin{figure}[!htbp]
\centering
\includegraphics[width=0.3\linewidth]{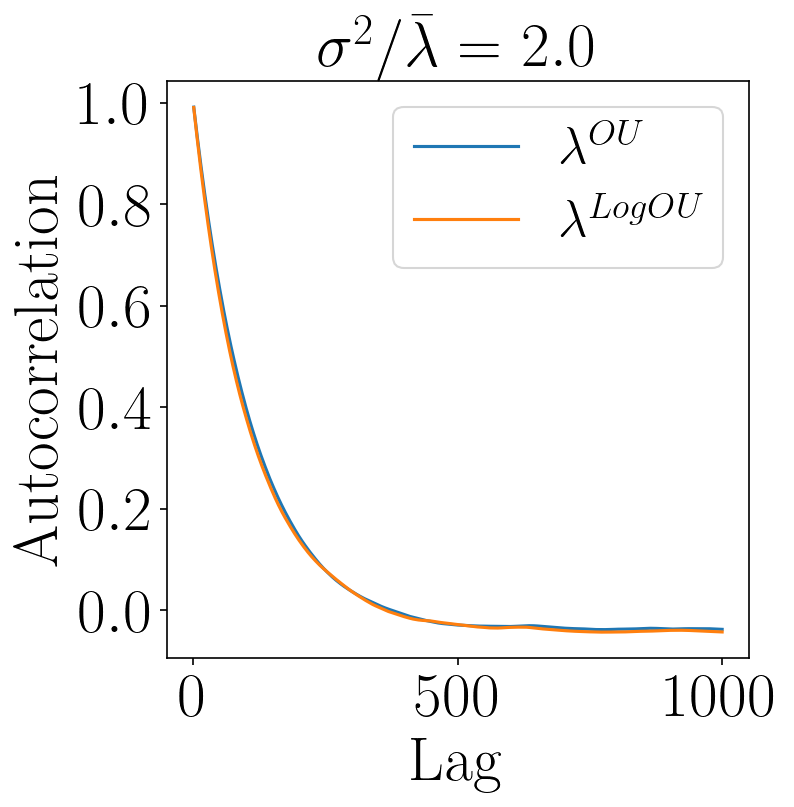}
\caption{Autocorrelation for contact rate $\lambda$ having $\bar{\lambda} = 1.4 \text{ per month}$ with white, OU and logarithmic OU noise perturbations---for a relative noise intensity of 2.0. Basic reproduction number $R_0 = 1.4$ (gonorrhea) was taken for these simulations.}
\end{figure}
\section{SIR Results with Correlation Time of 7 days}
\label{sec:appB}

\begin{figure}[!htbp]
\centering
\includegraphics[width = 0.4\linewidth]{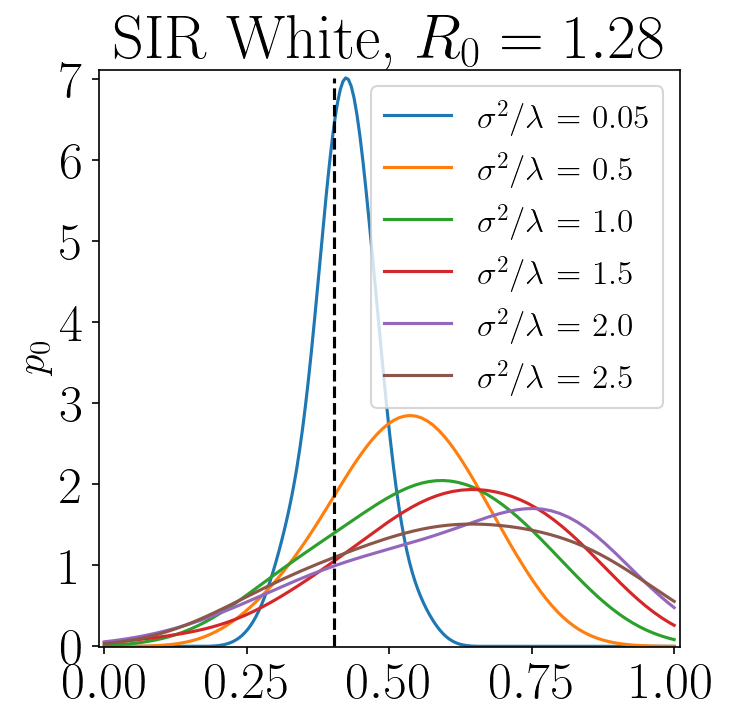}
\includegraphics[width = 0.4\linewidth]{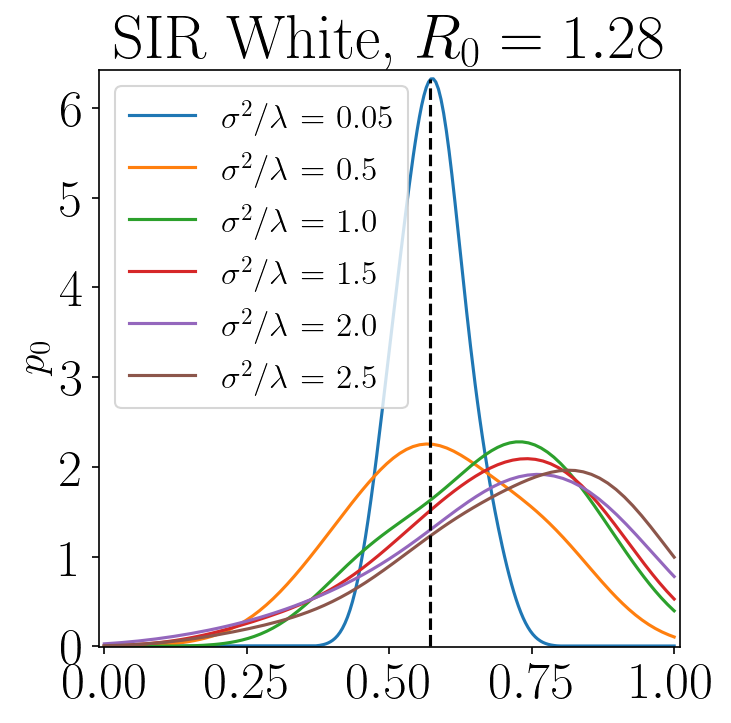}
\\
\includegraphics[width = 0.4\linewidth]{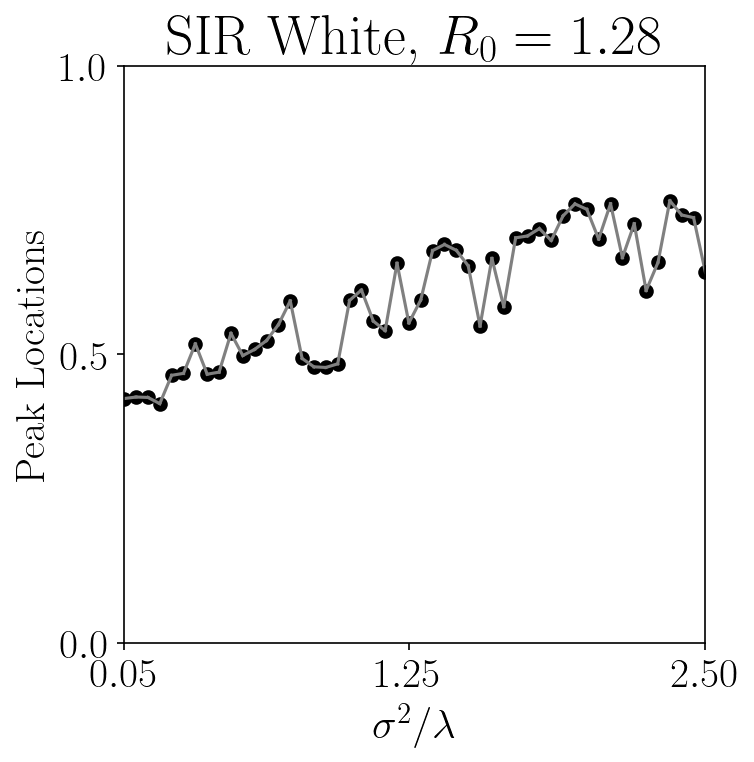}
\includegraphics[width = 0.4\linewidth]{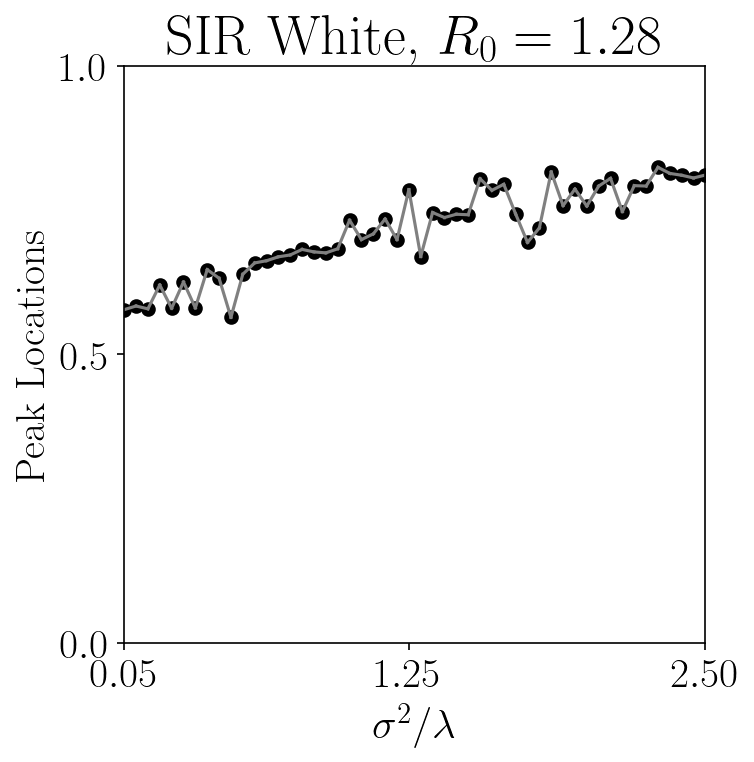}

\caption{Steady-state KDEs for SIR model over a range of relative noise intensities under white noise for flu with initial infected populations of $1\%$ (left) and $10\%$ (right), with corresponding peak-location-tracking plots.}

\label{fig:intensities_SIR_White_7}
\end{figure}

\begin{figure}[!htbp]
    \centering

\includegraphics[width = 0.4\linewidth]{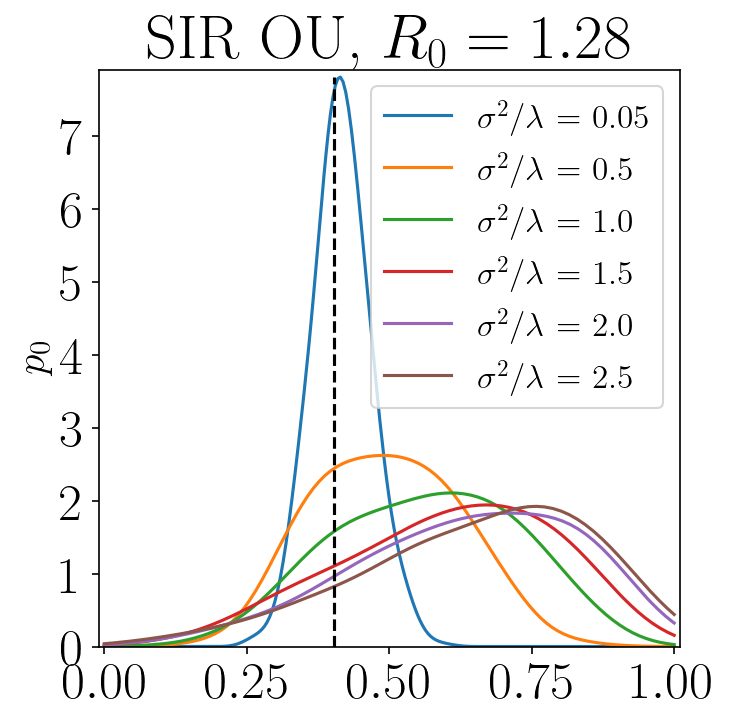}
\includegraphics[width = 0.4\linewidth]{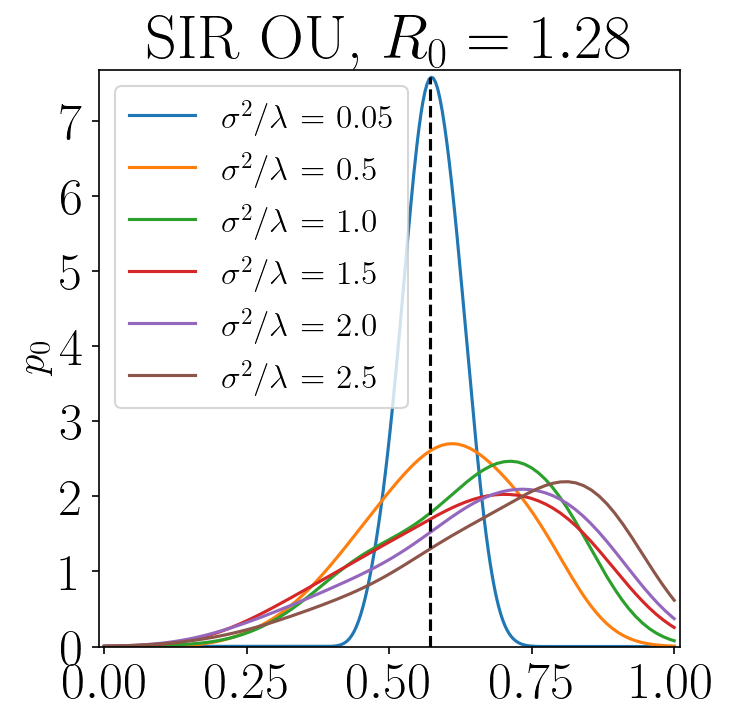}
\\
\includegraphics[width = 0.4\linewidth]{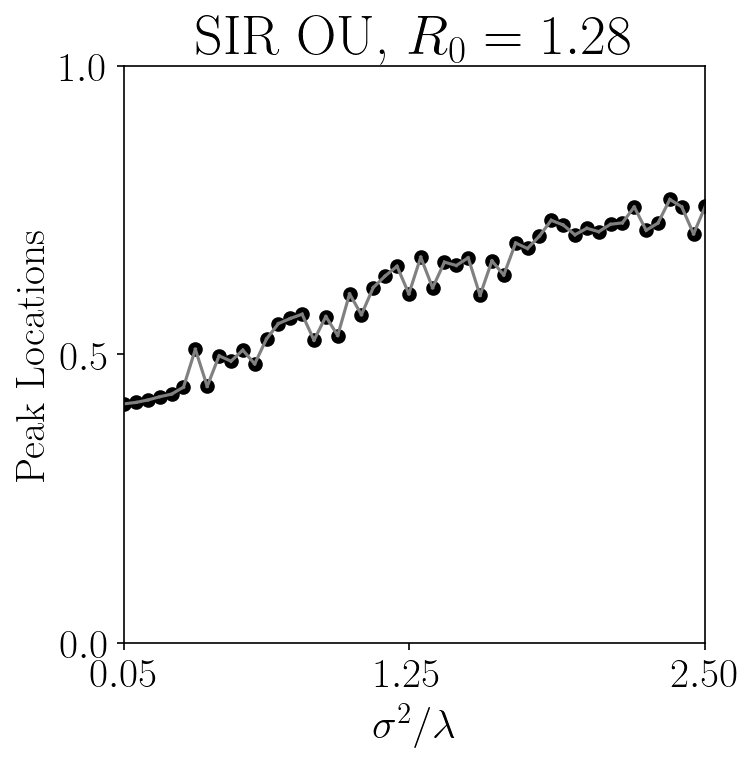}
\includegraphics[width = 0.4\linewidth]{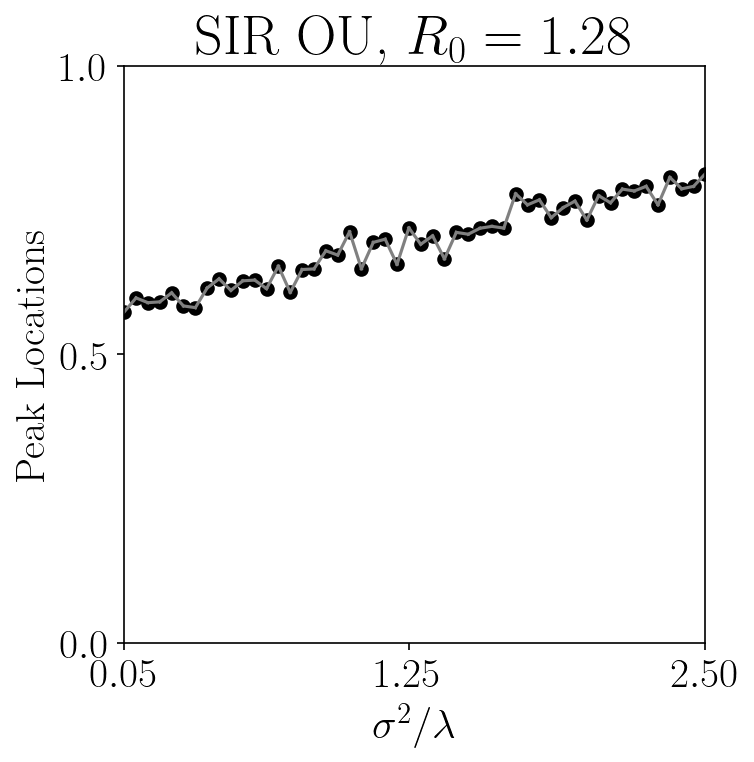}

\caption{Steady-state KDEs for SIR model over a range of relative noise intensities under OU noise for flu with $\tau = 7$ and initial infected populations of $1\%$ (left) and $10\%$ (right), with corresponding peak-location-tracking plots.}
    \label{fig:intensities_SIR_OU_7}
\end{figure}

\begin{figure}[!htbp]
    \centering
\includegraphics[width = 0.4\linewidth]{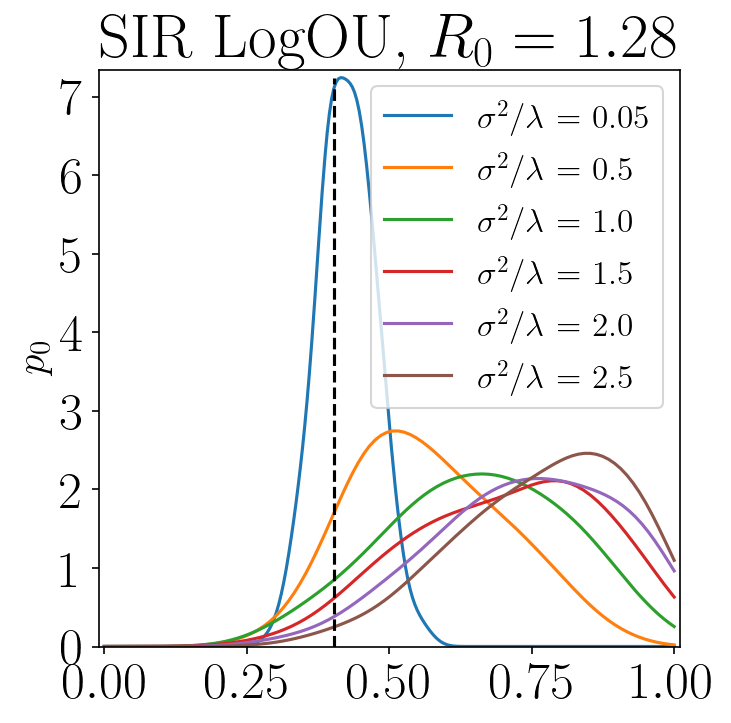}
\includegraphics[width = 0.4\linewidth]{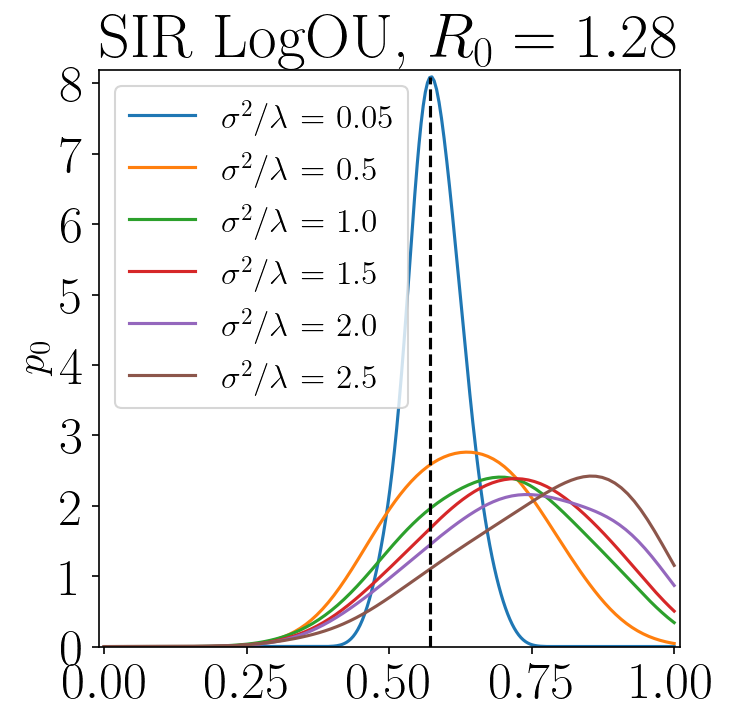}
\\
\includegraphics[width = 0.4\linewidth]{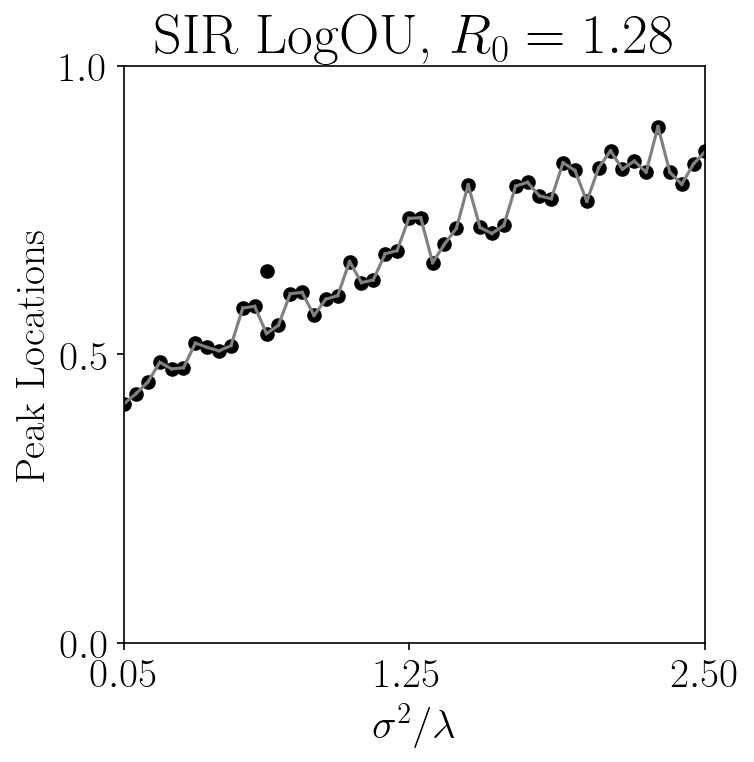}
\includegraphics[width = 0.4\linewidth]{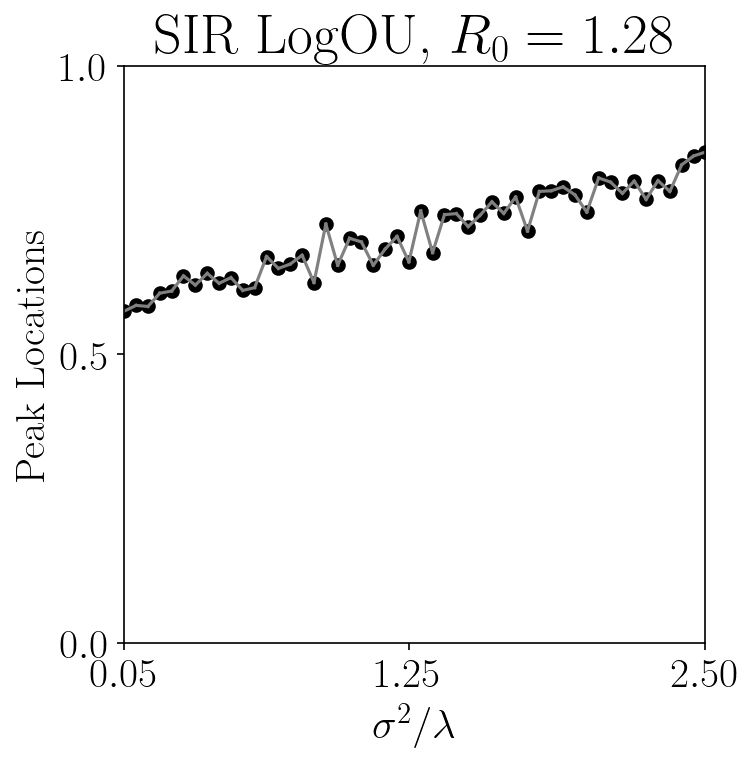}
    
\caption{Steady-state KDEs for SIR model over a range of relative noise intensities under LogOU noise for flu with $\tau = 7$ and initial infected populations of $1\%$ (left) and $10\%$ (right), with corresponding peak-location-tracking plots.}
\label{fig:intensities_SIR_LogOU_7}

\end{figure}

Similar to the results for SIR shown in Sec.~\ref{sec:R&D}, following Fig.~\ref{fig:overview_SIR_7} explores the KDEs of the removed fraction with a correlation time $\tau = 7$ days. The behaviour appears similar to the case of $\tau = 1$ except that the drift is now comparable for all noise types. Likewise, Figs.~\ref{fig:intensities_SIR_White_7},~\ref{fig:intensities_SIR_OU_7} and~\ref{fig:intensities_SIR_LogOU_7} show the KDEs with $\tau = 7$ days for various noise intensities with the corresponding peak-tracking plots.

\begin{figure}[!htbp]
\centering

\includegraphics[width = 0.3\linewidth]{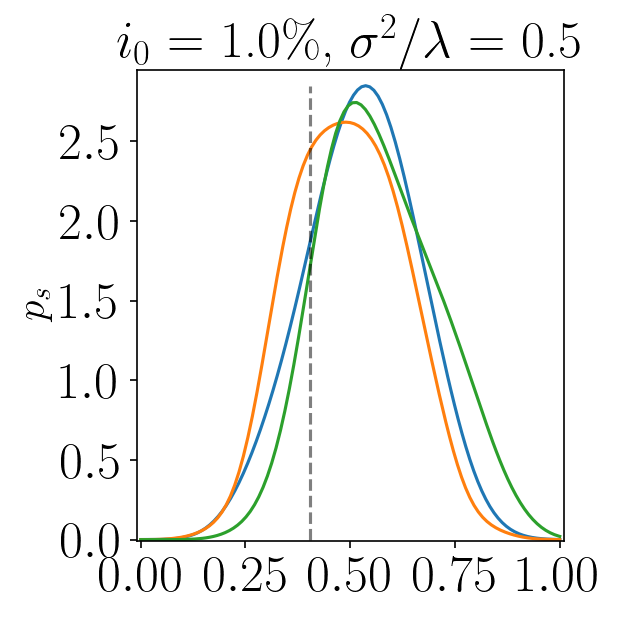}
\includegraphics[width = 0.3\linewidth]{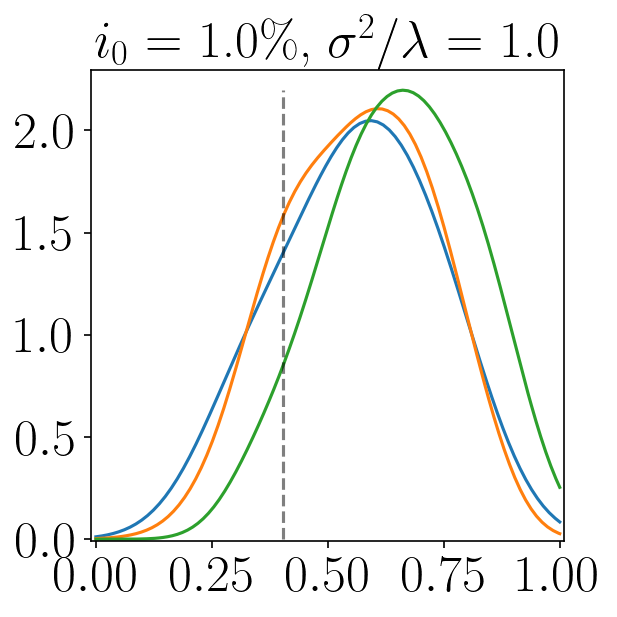}
\includegraphics[width = 0.3\linewidth]{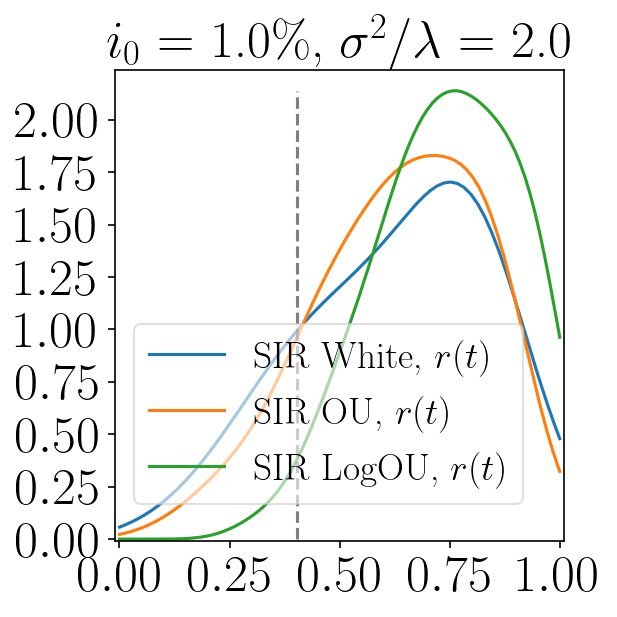}

\includegraphics[width = 0.3\linewidth]{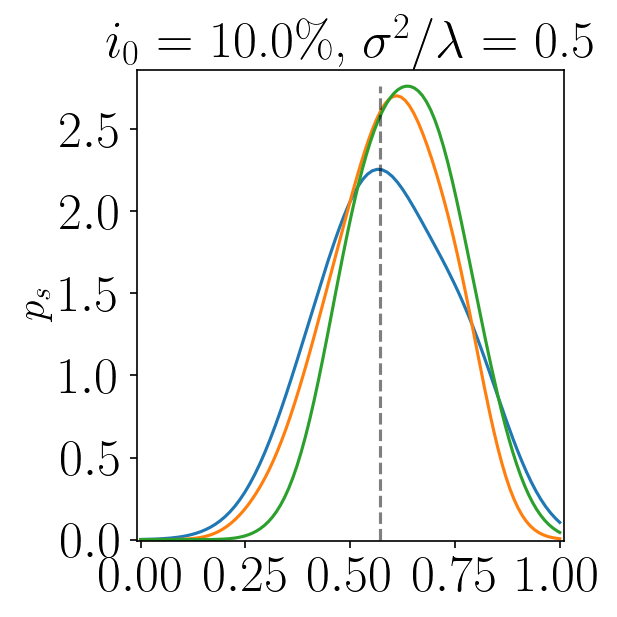}
\includegraphics[width = 0.3\linewidth]{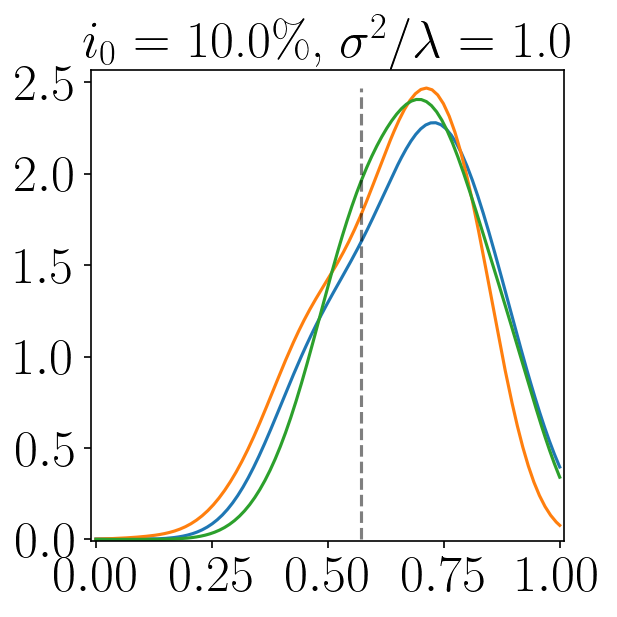}
\includegraphics[width = 0.3\linewidth]{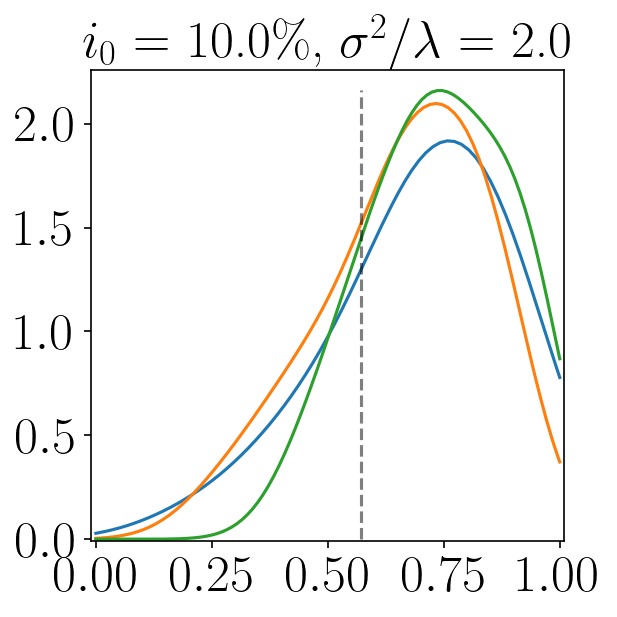}

\caption{Steady-state KDEs for flu ($R_0 = 1.28$) with $\tau = 7$ days.}
\label{fig:overview_SIR_7}

\end{figure}

\end{appendices}

\end{document}